\documentclass[useAMS,usenatbib,usegraphicx]{mn2e}
\usepackage{amsmath,amssymb}
\usepackage{color}
\def\gb {{\bf g}}

\def\rb {{\bf r}}
\def\ub {{\bf u}}
\def\xb {{\bf x}}
\def\nb {{\bf n}}

\newcommand{\Nyx}{\textsc{Nyx }}

\title[LES of isolated disc galaxies]{Large-eddy simulations of isolated disc galaxies with thermal and turbulent feedback}
\author[H.~Braun et al.]{H. Braun$^{1}$\thanks{E-mail:
hbraun@astro.physik.uni-goettingen.de}, W. Schmidt$^{1}$\thanks{E-mail:
schmidt@astro.physik.uni-goettingen.de}, J. C. Niemeyer$^{1}$, and A. S. Almgren$^{2}$\\
$^{1}$Institut f\"ur Astrophysik, Universit\"at G\"ottingen, Friedrich-Hund Platz 1, D-37077 G\"ottingen, Germany\\
$^{2}$Center for Computational Sciences and Engineering, Lawrence Berkeley National Laboratory, Berkeley, CA 94720, USA}

\begin{document}
\date{Accepted 2014 June 3. Received 2014 June 3; in original form 2013 December 18}
\pagerange{\pageref{firstpage}--\pageref{lastpage}} \pubyear{????}
\maketitle
\label{firstpage}
\begin{abstract}
We present a subgrid-scale model for the Multi-phase Interstellar medium, Star formation, and Turbulence (MIST) and explore its behaviour in high-resolution large-eddy simulations of isolated disc galaxies. 
MIST follows the evolution of a clumpy cold and a diffuse warm component of the gas within a volume element which exchange mass and energy via various cooling, heating and mixing processes. 
The star formation rate is dynamically computed from the state of the gas in the cold phase. 
An important feature of MIST is the treatment of unresolved turbulence in the two phases and its interaction with star formation and feedback by supernovae. 
This makes MIST a particularly suitable model for the interstellar medium in galaxy simulations. 
We carried out a suite of simulations varying fundamental parameters of our feedback implementation. 
Several observational properties of galactic star formation are reproduced in our simulations, such as an average star formation efficiency $\sim$1 per cent, a typical velocity dispersion around $\sim 10\ \mathrm{km\ s}^{-1}$ in star-forming regions, and an almost linear relationship between the column densities of star formation and dense molecular gas.
\end{abstract}
\section{Introduction}\label{sec:intro}
Stars are a product of a complex sequence of competing and interacting processes on a vast range of spatial and temporal scales that concentrate initially dilute gas into compact cores. 
It is not yet fully understood how the interplay of all of the processes involved, such as gravitational collapse, cooling, turbulence, magnetism, and stellar feedback, leads to the observed properties of the interstellar medium and stars in galaxies. 
A recent review on the properties of star formation was presented by \citet{Kennicut2012}. 
Of particular interest is the mechanism regulating the observed low efficiency of star formation. 
Measures of the star formation efficiency are gas depletion - or consumption - time-scales $\tau_{\rm dep}=M_{\rm gas}/\dot{M}_*$, which relate star formation to the available gas supply. 
\citet{WongBlitz2002,Evans2008,Bigiel2008,Blanc2009} infer $\tau_{\rm dep}\sim1-2$~Gyr in local disc galaxies from CO- $\mathrm{H\alpha}$-, and UV-measurements with resolutions down to 200~pc. 
Comparable measurements of gas-rich galaxies by \citet{Daddi2010,Tacconi2013,Saintonge2013} and others indicate a significantly shorter time $\tau_{\rm dep}\sim0.5$~Gyr,
corresponding to a relative gas consumption rate per free fall time $\epsilon_{\rm ff}\sim0.01$. 
According to the KS relation \citep[][and others]{Schmidt1959,Kennicutt1998}, the star formation rate is well correlated with the local gas supply. 
The power-law slope of measured KS relations depends, however, on the tracers used, the resolution achieved, and other observational limitations \citep[e.g.][]{Onodera2010,Lada2010,Leroy2013}. 
Recent observations show a good linear correlation between star formation rate and dense/molecular gas \citep{GaoSol2004,Lada2010,Bigiel2011}. \citet{Evans2009,Murray2011} showed that the local efficiency
$\epsilon_{\rm ff,MC}\simeq0.1$ in individual actively star-forming molecular clouds is much greater than the efficiency on galactic scales, 
and their depletion time-scale is considerably shorter ($\tau_{\rm dep}<100$~Myr). 
This implies that molecular clouds convert a sizable fraction 0.1-0.4 of their mass into stars during their lifetime \citep[a few 10~Myr, see e.g.][]{Blitz2007,McKee2007,Miura2012} before they are destroyed by supernova explosions (SNe) and stellar winds.\\

Actively star-forming molecular clouds are known to be strongly supersonically turbulent with typical velocity dispersions around $10\ \mathrm{km\ s}^{-1}$ \citep[e.g.][]{Leroy2008,Stilp2013}, or larger in interacting galaxies \citep[e.g.][]{Herrera2011}. 
Regulation by supersonic turbulence is a good candidate to theoretically explain the observed properties of star formation, as it globally supports a molecular cloud against gravity, but locally produces over-dense filaments and knots that may collapse into stars. 
A variety of approaches have been developed in the past years to derive star formation efficiencies from statistical properties of gravo-turbulent fragmentation inside molecular clouds \citep[][hereafter FK13]{PadNord09,Krumholz2005,HenChab11,Padoan2012,FedKless12}. 
As supersonic turbulence decays on relatively short time-scales of the order of the sound crossing time, it has to be maintained by some production mechanism over the lifetime of a molecular cloud. 
Processes such as large scale shear and instabilities in galactic discs \citep[e.g.][]{Gomez2002,Wada2002,Kim2003,Kim2007,AgerLake09,KrumBurk2010}, accretion of gas on to a galaxy \citep[e.g.][]{Hopkins2013,Genel2012,ElmeBurk09,KlessHenne10}, and merger events or other galactic interactions \citep[e.g.][]{Bournaud2011,Teyssier2010} come into question here, but also local processes like stellar winds \citep[e.g.][]{Wolf-Chase2000,Vink2000,Vink2011}, radiation pressure \citep{KrumThom2012} and SNe \citep[e.g.][]{OstrikerShetty2011,AgerLake09,Vollmer2003}, or the effects of thermal instabilities \citep[e.g.][]{WadaNorman2001,KritNor2002,Iwasaki2013} are possible turbulence production mechanisms.\\

In order to numerically simulate a realistic galaxy as a whole, star formation and the entailing stellar feedback have to be taken into account. 
However, the resolution to properly follow the evolution inside star-forming clouds in a galactic scale simulation is far from being feasible with contemporary computational resources. 
Recent simulations of isolated disc galaxies (IDG) feature resolutions down to a few parsec or even a fraction of a parsec \citep[e.g.][]{Hopkins2013,Renaud2013,Dobbs2013,Benincasa2013,Booth2013,Monaco2012}, while simulations of galaxies from cosmological initial conditions reach resolutions of some 10 parsec \citep[e.g.][]{AgerLake09,Munshi2013,Kraljic2012}. 
To tackle sub-resolution processes, an appropriate subgrid-scale (SGS) model has to be applied. 
In the last decade a wide range of different models have been devised to effectively describe star formation and stellar feedback \citep[e.g.][]{Agertz2013,StinSeth06,Stinson2013,Wise2012} using resolved quantities and assumptions about the small-scale properties of the ISM.\\

In galaxy simulations, the star formation rate $\dot{\rho_s}$ is usually modelled using a constant efficiency $\epsilon_{\rm ff}$ per free fall time, 
which locally enforces a KS relation
\begin{equation}
 \dot{\rho_s}=\epsilon_{\rm ff}\frac{\rho}{\tau_{\rm ff}}\propto\rho^{1.5},
\end{equation}
where $\rho$ is the local gas density and $\tau_{\rm ff}\propto\rho^{-0.5}$ the local free fall time. 
To avoid spurious star formation, additional constraints are applied, for example, a threshold for the minimal density required for star formation and a maximal temperature. 
More sophisticated models distinguish between different gas components.
\citet{GnedTass09} suggest to relate the star formation rate to the density of molecular gas instead of the total gas density. 
\citet{MurMona10} use a simple multi-phase approach to determine the fraction of the gas density that is available for star formation. 
For the simulations presented in this article, we use a multi-phase model for the ISM, including an estimation of the amount of shielded molecular gas \citep[][hereafter BS12]{BS12}. 
The star formation efficiency in the molecular gas is dynamically computed from the numerically unresolved turbulence energy, which is determined by a SGS model for compressible turbulence \citep[see ][hereafter SF11]{SchmFed10}. 
Since we incorporate the coupling between resolved and unresolved scales as turbulent stresses in the Euler equations, our galaxy simulations are large-eddy simulations (LES). 
Moreover, we apply the energy-conserving AMR techniques presented in \citet{Schmidt2013}. 
The diagonal part of the turbulent stresses acts as non-thermal pressure that usually dominates over the thermal pressure in cold and dense environments. 
This allows us to apply both thermal and turbulent feedback by channeling a fraction of the SN energy into the production of SGS turbulence energy. 
As we will show, this has important consequences for the regulation of star formation. 
In a way, this is similar to kinetic feedback \citep[see, e.g.,][]{Agertz2013}, with the important difference that we assume that turbulent motions are mainly excited on length scales below the grid resolution. 
For the thermal feedback, a small portion of the SNe energy is stored in a non-cooling budget - decaying on a time-scale of $1\ \mathrm{Myr}$ - to mimic the effect of hot SNe bubbles on sub-resolution scales, while the rest of the gas is allowed to cool radiatively. 
Although we include only effects of ionizing radiation from massive stars and SNe~II, we are able to reproduce several observational features of star formation and turbulence in quiescent, gas-rich (or high redshift $z\sim2$) disc galaxies and star-forming regions.\\

This paper is structured as follows. First we describe the numerical methods, the SGS model, and the setup of our isolated disc galaxy (IDG) simulations in Sections~\ref{sec:methods} and \ref{sec:simulations}. In Section~\ref{sec:result}, we present results from a suite of four simulations with different
treatments of feedback, followed by our conclusions in Section~\ref{sec:conclude}
\section{Numerical implementation}\label{sec:methods}
We carried out IDG simulations using the cosmological hydrodynamics code \Nyx \citep{NYX}. 
\textsc{Nyx}, built on the \textsc{BoxLib} software framework, uses Adaptive Mesh Refinement (AMR) to provide higher numerical resolution in sub-volumes of particular interest.
\Nyx solves the standard Euler equations using an unsplit Piecewise Parabolic Method (PPM); additional source terms are treated via a predictor/corrector scheme. 
\Nyx is capable of following the evolution of different collisionless massive particles in the $N$-Body formalism using a Kick--Drift--Kick algorithm, and (self-)gravity is taken into account using a Particle Mesh scheme with multigrid solver. 
We extended \Nyx to run adaptively refined LES of an IDG as described in the following.\\

To handle sub-resolution processes, such as star formation, stellar feedback, cooling, and thermal instability, we use a model based on BS12 with a few minor modifications, MIST (Multi-phase Interstellar medium, Star formation and Turbulence model). 
The key-features of MIST are the following.
\begin{itemize}
 \item Atomic, metal line, and dust cooling, photoelectric heating on dust.
 \item Separation of gas into two phases due to thermal instabilities that exchange energy and material via different mixing, heating, or cooling processes. The two phases represent a diffuse warm component and a clumpy, cold component of the gas. 
Balance of effective (i.e. thermal plus turbulent) pressure between the phases is assumed to obtain their respective densities. 
 \item Formation of stars from the molecular fraction of the cold phase that is shielded from dissociating radiation. 
The star formation rate is computed dynamically from the thermal and turbulent state of the gas.
 \item Depending on the age of a stellar population the stellar feedback is applied. 
Lyman-continuum radiation and SNe~II are taken into account. 
The SNe not only enrich the gas with metals, but also deposit kinetic energy and thermal energy into the gas. 
The SNe ejecta are treated as an additional sub-phase of the warm phase that does not  cool efficiently and is gradually mixed with the rest of the warm phase. 
 \item Unresolved turbulence is coupled to almost all processes implemented in MIST. 
Besides the source terms that are related to the scale separation for LES, small-scale pressure gradients caused by phase separation and SNe are taken into account.
\end{itemize}
An overview of important variables and coefficients is given in Table~\ref{tab:coeffs_n_params}.
\subsection{Gas dynamics}\label{sec:gasdyn}
As an extension to the standard compressible Euler equations we introduce a new degree of freedom in the form of the SGS turbulence energy density $\rho K$ (see SF11) and its source terms in order to model the behaviour of unresolved turbulent fluctuations. 
Furthermore we include the source terms as needed for the MIST model. The set of conservation equations for the evolution of gas becomes 
\begin{equation}\label{eq:dens}
 \frac{\partial \rho}{\partial t} + \nabla \cdot (\rho \ub) = -\dot{\rho}_{\rm s,SF}+\dot{\rho}_{\rm s,FB},
\end{equation}
\begin{equation}
\label{eq:momt}
\begin{split}
\frac{\partial (\rho \ub)}{\partial t} + \nabla \cdot \left[\rho \ub \ub + \left(p + \frac{2}{3}\rho K\right) - \tau^{\ast}\right] \\
   =\rho \gb -\ub\dot{\rho}_{\rm s,SF}+\ub_{\rm s,FB}\dot{\rho}_{\rm s,FB},
\end{split}
\end{equation}
\begin{equation}
\label{eq:energy}
\begin{split}
\frac{\partial (\rho E)}{\partial t} + \nabla \cdot \left[\rho \ub E + \left(p + \frac{2}{3}\rho K\right)\ub - \ub\cdot\tau^{\ast}\right]  \\ 
 =\rho \ub \cdot \gb -\Lambda - \Pi_{\rm TI} - \Pi_{\rm SGS} -(E-e+e_{\rm c})\dot{\rho}_{\rm s,SF}\\
  +\left(e_{\rm SN}+\frac{\ub\left(2\ub_{\rm s,FB}-\ub\right)}{2}\right)\dot{\rho}_{\rm s,FB}-\Pi_{\rm SN}, \\
\end{split}
\end{equation}
\begin{equation}
 \label{eq:turb}
\begin{split}
\frac{\partial (\rho K)}{\partial t} + \nabla \cdot \left(\rho \ub K - \rho \kappa_{\rm SGS} \nabla K\right)  \\
  =\Pi_{\rm SGS} - \rho \varepsilon_{\rm SGS}+\Pi_{\rm TI}-K\dot{\rho}_{\rm s,SF}\\
  +\Pi_{\rm SN} + \dot{\rho}_{\rm s,FB}^{\rm ex}.
\end{split}
\end{equation}
Here $\rho$ is the gas density, $\ub$ the velocity vector, $E$ the total specific energy of the gas, $K$ the specific turbulent SGS-energy, $e$ the specific internal energy, $p=(\gamma-1)\rho e$ the thermal gas pressure (where $\gamma=5/3$ is the polytropic equation of state parameter), and $\gb$ the gravitational acceleration vector (see Section~\ref{sec:grav}).
The SGS turbulence model related quantities $\Pi_{\rm SGS}$, $\rho\varepsilon_{\rm SGS}$, and $\kappa_{\rm SGS}$ are defined in Section~\ref{sec:sgs_mod}. For a definition of the star formation rate $\dot{\rho}_{\rm s,SF}$ see Section~\ref{sec:sfr}. 
The stellar feedback rate $\dot{\rho}_{\rm s,FB}$, the mass-weighted average velocity $\ub_{\rm s,FB}$ of SN-ejecta, and the specific supernova energy deposit $e_{\rm SN}=6.5\times10^{49}\ \mathrm{erg}/M_\odot$ are explained in Section~\ref{sec:fb_hot_phase}. 
We apply a multiphase model to determine the specific thermal energy $e_{\rm c}(T_{\rm c}=50\ \mathrm{K})$ of the cold phase, and the net 
cooling rate $\Lambda$ (see Section~\ref{sec:ti_phases}) and the turbulence energy production via phase separation $\Pi_{\rm TI}$ and via SN feedback energy deposit $\Pi_{\rm SN}$ (see Section~\ref{sec:sfr}). 
To conserve the kinetic energy of SN-ejecta, $\dot{\rho}_{\rm s,FB}^{\rm ex}$ is added to $\rho K$ (see equation~\ref{eq:excess} in Section~\ref{sec:fb_hot_phase}).

 \begin{table*}
 \caption{Important variables and coefficients.}
 \begin{minipage}{177mm}
 \label{tab:coeffs_n_params}
 \centering
 \begin{tabular}{llll}
   Symbol&Value&Description&Reference\\
 \hline
 \hline
 \multicolumn{4}{|c|}{Hydrodynamics} \\
 \hline
 $\rho$&&Total gas density&Equation~(\ref{eq:dens})\\
 $\rho\ub$&&Vector of linear momentum density of gas&Equation~(\ref{eq:momt})\\
 $\rho E$&&Thermal plus (resolved) kinetic energy density of gas&Equation~(\ref{eq:energy})\\
 $\rho K$&&SGS turbulence energy density&Equation~(\ref{eq:turb})\\
 $\rho_{\rm c}$&&Fractional density of cold phase gas&BS12, equation~(\ref{eq:cdens})\\
 $\rho_{\rm h}$&&Fractional density of hot SNe ejecta&Equation~(\ref{eq:hdens})\\
 $\rho_{\rm H}$&&Hydrogen density&Equation~(\ref{eq:specdens})\\
 $\rho_{\rm He}$&&Helium density&Equation~(\ref{eq:specdens})\\
 $\rho_{\rm Z}$&&Metal density&Equation~(\ref{eq:specdens})\\
 $\gamma$&$5/3$&Polytropic index of equation of state&\\
 \hline
 \multicolumn{4}{|c|}{SGS turbulence} \\
 \hline
 $\Pi_{\rm SGS}$&&Production rate of SGS energy by turbulent stresses&SF11, equation~(\ref{eq:pi_casc})\\
 $\tau$&&Turbulent stress tensor&SF11, equation~(\ref{eq:turb_stresses})\\
 $C_{\rm \tau1}$&0.02&Linear closure coefficient of turb. stresses&SF11\\
 $C_{\rm \tau2}$&0.75&Non-linear closure coefficient of turb. stresses&SF11\\
 $C_{\rm \varepsilon}$&1.58&Dissipation coefficient of SGS turb. energy&SF11\\
 $C_{\rm \kappa}$&0.65&Diffusion coefficient of SGS turb. energy&SF11\\
 \hline
 \multicolumn{4}{|c|}{MIST} \\
 \hline
 $\rho_{\rm c,pa}$&&Average density of cold phase gas&BS12\\
 $\rho_{\rm w,pa}$&&Average density of warm phase gas&BS12\\
 $e_{\rm w}$&&Specific thermal energy of warm gas&BS12\\
 $\ell_{\rm c}$&&Length scale of cold phase clumps&BS12\\
 $\dot{\rho}_{\rm s,SF}$&&Star formation density&BS12, equation~\ref{eq:sfrMIST}\\
 $\dot{\rho}_{\rm s,FB}$&&SNe feedback density&BS12, Sec.~\ref{sec:fb_hot_phase}\\
 $\epsilon_{\rm PN}$&&Star formation free-fall time efficiency in cold gas&PN11, FK13, equation~(\ref{eq:eps_PN})\\
 $\Pi_{\rm SN}$&&Production rate of SGS energy via SNe feedback&Equation~\ref{eq:pi_sn}\\
 $\Pi_{\rm TI}$&&Production rate of SGS energy via phase separation&Equation~\ref{eq:pi_ti}\\
 $A_{\rm SN}$&&SNe evaporation coefficient&BS12, equation~(\ref{eq:A_SN})\\
 $f_{\rm TI}$&0 or 1&Thermal instability switch&BS12\\
 $\epsilon_{\rm TI}$&0.025&Efficiency of SGS energy production by phase separation&BS12\\
 $\epsilon_{\rm SN}$&0.085&Efficiency of SGS energy production by SNe feedback&BS12\\
 $e_{\rm SN}$&$6\times10^{49}\ \mathrm{erg}\ \mathrm{M_\odot^{-1}}$&Energy release per $\mathrm{M_\odot}$ of SNe~II&BS12\\
 $\tau_{h}$&$1\ \mathrm{Myr}$&Decay parameter of the hot SNe gas&Sec.~\ref{sec:fb_hot_phase}\\
 $e_{\rm h}$&$0.1\times e_{\rm SN}$&Specific thermal energy of gas in SNe bubbles&Sec.~\ref{sec:fb_hot_phase}\\
 $e_{\rm c}$&$e(T_{\rm c}\equiv50\mathrm{\ K})$&Specific thermal energy of cold gas&BS12\\
 $f_{\rm loss}$&0.4&Fraction of prestellar mass loss&BS12\\
 $b$&$1/3\ldots1$&Compressive factor, density PDF broadening parameter&BS12, equation~(\ref{eq:b_comp})\\
 $\eta$&$1/3$&Turbulent velocity scaling coefficient of warm gas&BS12\\
 $\zeta$&0.1&Metal loading fraction of SNe ejecta&BS12\\
  \hline
 \multicolumn{4}{l}{\textbf{Abbreviated references}: BS12 - \citet{BS12}, FK13 - \citet{FedKless12},}\\
 \multicolumn{4}{l}{PN11 - \citet{PadNord09}, SF11 -\citet{SchmFed10}.}
 \end{tabular}
 \end{minipage}
 \end{table*}
\subsubsection{Gravity}\label{sec:grav}
The massive components in our IDG simulations are dark matter, baryonic gas, and stars, where the dark matter component is assumed to be a static halo, and the other two are dynamically evolved. 
The gravitational acceleration vector $\gb$ is computed as the sum of a static acceleration due to the dark matter halo and the negative gradient of the gravitational potential due to the dynamical components:
\begin{equation}
 \gb = \gb_{\rm dm} - \nabla\Phi_{\rm dyn},
\end{equation}
 with the static acceleration $\gb_{\rm dm}$ (see equation~\ref{eq:dm_accel} in Section~\ref{sec:initial_cond}). $\Phi_{\rm dyn}$ represents the solution of Poisson's equation
\begin{equation}
 \bigtriangleup\Phi_{\rm dyn}=4\pi G(\rho_{\rm dyn}-\bar{\rho}_{\rm dyn}),
\end{equation}
where $\rho_{\rm dyn}=\rho+\rho_{\rm s}$, $\bar{\rho}_{\rm dyn}$ is the mean of $\rho_{\rm dyn}$, $\rho_{\rm s}$ is the total stellar density, and $G$ is the gravitational constant.
\subsubsection{Hydrodynamical turbulence model}\label{sec:sgs_mod}
The interaction between resolved and unresolved turbulent velocity fluctuations is modelled using the SGS turbulence stress tensor $\tau$, which can be seen as an analogue to the viscous dissipation tensor in the Navier-Stokes equations. 
With $u_{i,k}:=\partial u_i/\partial x_k$ its components following SF11 read
\begin{equation}
\label{eq:turb_stresses}
\begin{split}
\tau_{ij} = 2C_{\tau1}\Delta(2\rho K)^{1/2}S_{ij}^{\ast} - 2C_{\tau2}\rho K\frac{u_{i,k}u_{j,k}}{u_{l,m}u_{l,m}} \\
  - \frac{2}{3}(1-C_{\tau2})\rho K\delta_{ij},\\
\end{split}
\end{equation}
where
\begin{equation}
S_{ij}^{\ast} = S_{ij} - \frac{1}{3}\delta_{ij}d =
\frac{1}{2} (u_{i,j} + u_{j,i}) - \frac{1}{3}\delta_{ij}u_{k,k}
\end{equation}
is the trace-free rate of strain, and $\Delta$ the grid scale. 
The trace free stress tensor $\tau^{\ast}$, used in equations~(\ref{eq:momt}) and (\ref{eq:energy}), is given by $\tau^{\ast}_{\rm ij}=\tau_{\rm ij} - 2\delta_{\rm ij}K/3$.
The SGS turbulence energy production rate $\Pi_{\rm SGS}$, the SGS turbulence dissipation rate $\rho\varepsilon_{\rm SGS}$ (which does not appear in equation~\ref{eq:energy} as it is absorbed into $\Lambda$), and the SGS turbulence diffusivity $\kappa_{\rm SGS}$ are given by
\begin{eqnarray}
\Pi_{\rm SGS}             &=&\tau_{ij} S_{ij}, \label{eq:pi_casc}\\
(\rho \varepsilon_{\rm SGS}) &=& \frac{\rho C_\varepsilon K^{3/2}}{\Delta},\\
\kappa_{\rm SGS}   &=& C_\kappa(2K)^{1/2}.
\end{eqnarray}
We use the closure coefficients $C_{\tau1}=0.02$, $C_{\tau2}=0.75$, $C_\varepsilon=1.58$, and $C_\kappa=0.65$ as determined by SF11 for compressible turbulence.
\subsection{Non-adiabatic physics}
In the following we describe how unresolved physics such as heating, cooling, star formation and stellar feedback was implemented in the code \textsc{Nyx}. 
For a more detailed description of the underlying model we refer to BS12.
As input for the computation of the non-adiabatic physics sources we need the hydrodynamical state, the source terms belonging only to the SGS turbulence model and the stellar feedback terms. 
Contrary to BS12, stellar feedback is considered an external source in the calculation, as it depends on the stellar population represented by $N$-body particles but not on the hydrodynamical state. 
Given the SGS- and stellar feedback source terms, the actual sources are calculated by subcycling the BS12 model ODEs locally in a grid cell, to resolve all time-scales of relevant processes, particularly the cooling time-scale, and then averaging the rate of change over the hydro-step. 
To follow the metal enrichment, we calculate three hydrogen density $\rho_{\rm H}$, helium density $\rho_{\rm He}$ and metal density $\rho_{\rm Z}$. Their conservation equations are of the form
\begin{equation}\label{eq:specdens}
 \frac{\partial\rho_{\rm X}}{\partial t} + \nabla \cdot (\rho_{\rm X} \ub) =-\frac{\rho_{\rm X}}{\rho}\dot{\rho}_{\rm s,SF}-\left.\frac{\partial\rho_{\rm s,X}}{\partial t}\right|_{\rm FB},
\end{equation}
where $X$ indicates one of the species H, He, or Z, and $\left.\partial\rho_{\rm s,X}/\partial t\right|_{\rm FB}=:\dot{\rho}_{Xs,FB}$ is the ejection rate of that species by SNe (see Section~\ref{sec:fb_hot_phase}).
\subsubsection{Cold and warm gas phases}\label{sec:ti_phases}
To keep track of the multiphase state in a grid cell, we introduce an additional passively advected quantity, the cold-phase fractional density $\rho_{\rm c}$, from which we can easily reconstruct the warm-phase density $\rho_{\rm w}=\rho-\rho_{\rm c}$. 
The warm phase thermal energy is given by $\rho_{\rm w}e_{\rm w}=\rho e-\rho_{\rm c}e_{\rm c}$ with a constant specific thermal energy $e_{\rm c}$ of the cold phase, corresponding to a temperature $T_{\rm c}=50\ \mathrm{K}$. The conservation equation of $\rho_{\rm c}$ reads
\begin{equation}\label{eq:cdens}\begin{split}
 \frac{\partial\rho_{\rm c}}{\partial t} + \nabla \cdot (\rho_{\rm c} \ub) =& \frac{\Lambda_{\rm c}+\Lambda_{\rm w}f_{\rm TI}}{e_{\rm w}-e_{\rm c}}\\
 &-\dot{\rho}_{\rm s,SF}-A_{\rm SN}\dot{\rho}_{\rm s,FB},
\end{split}\end{equation}
where 
\begin{align*}
 \Lambda_{\rm c}=-\rho_{\rm c}\varepsilon_{\rm SGS}-\Gamma_{\rm c}^{\rm PAH}-\frac{\rho_{\rm c}}{\rho}\Gamma^{\rm Lyc}\mbox{ and }\\
 \Lambda_{\rm w}=\Lambda_{\rm w}^{\rm rad}-\rho_{\rm w}\varepsilon_{\rm SGS}-\Gamma_{\rm w}^{\rm PAH}-\frac{\rho_{\rm w}}{\rho}\Gamma^{\rm Lyc}
\end{align*}
are the net cooling rates of the cool and warm phase, respectively. $\Lambda_{\rm c}$ is effectively a heating rate. Material is removed from the cold phase and transferred to the warm phase instead of increasing $u_{\rm c}\equiv \mathrm{const}$.
\begin{equation}\label{eq:A_SN}
 A_{\rm SN}=13826\left(\dfrac{\rho_{\rm w,pa}}{\frac{\mathrm{m_H}}{\mathrm{cm^{3}}}}\right)^{\frac{4}{5}}\left(\dfrac{\ell_{\rm c}}{\mathrm{pc}}\right)^{-\frac{6}{5}}\left(\dfrac{\rho_{\rm c}}{\rho_{\rm c,pa}}\right)^\frac{3}{5}
\end{equation}
 is the SN cold-phase evaporation coefficient, and $\epsilon_{\rm TI}=0.025$ is the efficiency parameter for turbulence production by the thermal instability, if the indicator $f_{\rm TI}=1$ (see BS12). 
Here $\Gamma^{\rm PAH}$ is the photoelectric heating rate, $\Gamma^{\rm Lyc}$ the heating rate due to Lyman continuum radiation from young, massive stars (see Section~\ref{sec:fb_hot_phase}), and
\begin{equation}\label{eq:lambw}
\Lambda_{\rm w}^{\rm rad}=\frac{\hat{\rho}_{\rm w}}{\rho_{\rm w,pa}}\hat{\Lambda}_{\rm w}^{\rm rad}(\rho_{\rm w,pa},Z,\hat{T}_{\rm w})
\end{equation}
the radiative cooling rate, which is interpolated from a cooling table. These tabled cooling rates were computed using the photo-ionization program package \textsc{Cloudy} \citep[version 08.00]{Ferland1998}. 
$\rho_{\rm w,pa}$ and $\rho_{\rm c,pa}$ are the average densities of the warm and the cold phase, respectively. 
Those are computed from the fractional phase densities ($\rho_{\rm c}$ and $\rho_{\rm w}$) and the thermal and turbulent energies ($e_{\rm c}$, $e_{\rm w}$, and $K$) by assuming balance of effective (thermal plus turbulent) pressure between the phases at cold clump scale $\ell_{\rm c}$, as explained in detail in BS12.
$Z$ is the metallicity of the gas. $\hat{T}_{\rm w}=\hat{T}_{\rm w}(\hat{e}_{\rm w},Z)$ and $\hat{\rho}_{\rm w}$ are the temperature and the fractional density of the warm-phase gas, that is allowed to cool radiatively. 
Note that $\hat{T}_{\rm w}$ and $\hat{\rho}_{\rm w}$ may differ from $T_{\rm w}$ and $\rho_{\rm w}$ in areas affected by recent SNe feedback. 
The treatment of the third gaseous phase, the hot SNe ejecta, which is prevented from cooling, is described in Section~\ref{sec:fb_hot_phase}. 
The total net cooling rate, used in equation~(\ref{eq:energy}), is then given by $\Lambda=\Lambda_{\rm c}+\Lambda_{\rm w}$.
\subsubsection{Star formation rate}\label{sec:sfr}
Stars are assumed to form from the molecular fraction of the gas in the cold phase, $f_{\rm H_2}\rho_{\rm c}$ at a rate \citep{Krumholz2009}
\begin{equation}\label{eq:sfrMIST}
 \dot{\rho}_{\rm s,SF}=\frac{f_{\rm H_2}\rho_{\rm c}\epsilon_{\rm PN}}{t_{\rm c,ff}},
\end{equation}
where $\epsilon_{\rm PN}$ is the formation rate of gravitationally bound cores per free fall time
\begin{equation}
 t_{\rm c,ff}=\sqrt{\frac{3\pi}{32G\rho_{\rm{c,pa}}}}.
\end{equation}
To calculate $\epsilon_{\rm PN}$, we use the \citet[hereafter PN11]{PadNord09} model in the single free-fall formulation of FK13
\begin{equation}\label{eq:eps_PN}
 \epsilon_{\rm PN}=\frac{(1-f_{\rm loss})r_{\rm crit}^{\frac{1}{2}}}{2}\left(1+\mathrm{erf}\left[\frac{\sigma_{\rm c}^2-2\log\left(r_{\rm crit}\right)}{\left(8\sigma_{\rm c}^2\right)^{\frac{1}{2}}}\right]\right).
\end{equation}
Here $f_{\rm loss}=0.4$ is the fraction of mass in gravitationally bound cores lost during prestellar collapse through winds etc.,  
and $\sigma_{\rm c}=\sqrt{\log\left(1+b^2\mathcal{M}^2_{\rm c}\right)}$ the standard deviation of the assumed density probability density function (PDF) of log-normal shape. 
The broadening parameter $b$ is set depending on which of the three production terms of turbulence energy in equation~(\ref{eq:turb}) is locally the dominant one
\begin{equation}
 \Pi_{\rm max} =\max\left[\Pi_{\rm SN},\Pi_{\rm TI},\Pi_{\rm SGS}\right],
\end{equation}
 where
\begin{equation}\label{eq:pi_ti}
 \Pi_{\rm TI}=\Lambda_{\rm w}f_{\rm TI}\epsilon_{\rm TI}
\end{equation}
is the contribution due to thermal instability, and
\begin{equation}\label{eq:pi_sn}
 \Pi_{\rm SN}=\dot{\rho}_{\rm s,FB}e_{\rm SN}\epsilon_{\rm SN}
\end{equation}
describes turbulence production by SNe. $\epsilon_{\rm SN}=0.085$ is the fraction of the energy released by SN that is deposited in the form of turbulent energy. 
We define $b$ by
\begin{equation}\label{eq:b_comp}
  b =
  \begin{cases}
     1/3 & \mbox{ if }\Pi_{\rm SGS}=\Pi_{\rm max} \\
     2/3 & \mbox{ if }\Pi_{\rm TI}=\Pi_{\rm max}\\
     1 & \mbox{   if }\Pi_{\rm SN}=\Pi_{\rm max} .
  \end{cases}
\end{equation}
Here we assume the large-scale driving $\Pi_{\rm SGS}$ to be mostly caused by shear, the SNe driving to be mostly compressive, and the thermal instability driving to be of intermediate type. The corresponding values follow from \citet{Federrath2010b}.\\

To obtain the turbulent Mach-number $\mathcal{M}_{\rm c}$ of the cold phase, the SGS energy $K$ has to be rescaled from the grid scale $\Delta$ to the cold clump scale $\ell_{\rm c}$, assuming a Kolmogorov velocity scaling exponent $\eta=1/3$:
\footnote{We assume $\ell_{\rm c}$ to be the largest scale represented in the cold phase. Consequently, the scaling of the turbulent velocities is applied to those scales, on which only the warm phase exists. 
Turbulence in the warm gas is usually subsonic or transonic at most. 
In this regime the assumption of a Kolmogorov-type scaling behaviour with coefficient $\eta=1/3$ seems valid.}
\begin{equation}
 \mathcal{M}_{\rm c}^2 = \frac{2K\left(\frac{\ell_{\rm c}}{\Delta}\right)^{2\eta}}{\gamma(\gamma-1)e_{\rm c}}.
\end{equation}
The critical over-density ratio $r_{\rm crit}=\rho_{\rm crit}/\rho_{\rm c,pa}$, above which a bound object is formed, is given by (FK13)
\begin{equation}
 r_{\rm crit}=0.0067\times\frac{5\times2K\left(\frac{\ell_{\rm c}}{\Delta}\right)^{2\eta}}{\pi G\rho_{\rm c,pa}\ell_{\rm c}^2}\mathcal{M}_{\rm c}^2.
\end{equation}

The molecular fraction of cold gas $f_{\rm H_2}$ is computed from the cold and warm phase fractional densities and energies, assuming effective pressure equilibrium, using a St\"omgren-like approach. 
The penetration depth of impinging radiation into a spherical cold clump of diameter $\ell_{\rm c}$ is determined by the balance between $\mathrm{H}_2$ production and dissociation due to UV-photons. 
The dissociating radiation field $I_{\nu}$ is assumed to be homogenous and isotropic. However, in dense, cold environments, which are identified by $\rho_{\rm c,pa}>10\times \mu m_{\rm H}\ \mathrm{cm}^{-3}$ and $T(e,Z)<1000\;\mathrm{K}$, $I_{\nu}$ is dimmed by a factor 
\begin{equation}
 f_{I_\nu}=\frac{1}{3}+\frac{2}{3}\max\left[1+\frac{T_{I_\nu}-T(e,Z)}{T_{I_\nu}},0\right],
\end{equation}
with $T_{I_\nu}=500\;\mathrm{K}$, because of the assumed shielding from radiation by the environment.\\

The assumption of a log-normal shaped density PDF, which is an essential part of the theory of PN11, applies to turbulence in isothermal gas.
For a consistent definition of the internal energy of the cold phase in MIST, an adiabatic exponent $\gamma=5/3$ is required. 
However, we assume a constant average temperature of the cold phase because of processes which are not explicitly treated.
Both observational and numerical studies on the density PDFs show that the density PDF of the cold phase of the ISM is well approximated by a log-normal PDF \citep[e.g.][]{Hughes2013,Schneider14}.
Although a power-law tail is generally found for actively star-forming clouds in which dense cores undergo gravitational collapse, FK13 point out that this does not significantly affect the star formation efficiency because the log-normal turbulent density fluctuations feed the collapsing gas that populates the power-law tail at high densities. Despite of the underlying inconsistency, all currently available analytic models for the calculation of star formation efficiencies, including PN11, are based on this assumption. Substituting PDFs with power-law tails into these models does not amend the problem because this would lead to divergent integrals. As a consequence, the construction of consistent models of the star formation efficiency is an open problem. 
\subsection{Implementation of stellar particles}
\subsubsection{Stellar particle creation}\label{sec:spart_creation}
A particle is characterized by its position $\xb_{\rm p}$, mass $m_{\rm p}$, velocity $\ub_{\rm p}$, and an arbitrary number of additional properties.
To handle the dynamical evolution of stars, we implemented a particle type with three additional properties: 
the initial mass $m_{\rm pi}$, creation time $t_{\rm pc}$, and metallicity $Z_{\rm p}$, which are needed for the application of stellar feedback. 
A stellar particle does not represent a single star, but a single stellar population with a normalized initial mass function (IMF) $\mathrm{d}N_{\rm *}/\mathrm{d}m_{\rm *}$ (where $N_{\rm *}$ is the number of stars of individual initial mass $m_{\rm *}$ per solar mass of stellar population) scaled by $m_{\rm pi}$. 
We use the IMF of \cite{Chabrier2001}.\\
To avoid the repeated creation of particles in all cells where $dot{\rho}_{\rm s,SF} > 0$, we introduce a stellar density field $\rho_{\rm s,m}$, that acts as an intermediate buffer for the stellar mass. 
This is treated as an passively advected quantity with respect to the hydro-solver, but massive with respect to gravity. Its conservation equation reads
\begin{equation}\begin{split}
 \frac{\partial\rho_{\rm s,m}}{\partial t} + \nabla \cdot (\rho_{\rm s,m} \ub) =&\dot{\rho}_{\rm s,SF}-\left.\frac{\partial\rho_{\rm s,m}}{\partial t}\right|_{\rm SP},
\end{split}\end{equation}
where $\left.\partial\rho_{\rm s,m}/\partial t\right|_{\rm SP}$ represents the mass transfer from $\rho_{\rm s,m}$ into stellar particles. The total stellar mass density is given by $\rho_{\rm s}=\rho_{\rm s,p}+\rho_{\rm s,m}$.\\
A pair of stellar particles $p^1$ and $p^2$ is created in the cell centre, if the agglomerated $\rho_{\rm s,m}$ exceeds the threshold $\rho_{\rm s,m,max}\propto\Delta^{-2}$ (corresponding to a minimum particle pair mass $2 m_{\rm p,min}=\rho_{\rm s,m,max}\Delta^{-3}\approx 40\;\mathrm{M}_\odot$ for a cell size of $\Delta\approx 30\;\mathrm{pc}$), the mass is removed from $\rho_{\rm s,m}$. The properties of the new particles are
\begin{equation}\begin{split}
 m_{\rm p}^{1,2} = &\; \rho_{\rm s,m}\Delta^{3}/2,\\
 \ub_{\rm p}^{1,2} = &\; \ub \pm \ub_{\rm rnd},\\
 m_{\rm pi}^{1,2} = &\; \rho_{\rm s,m}\Delta^{3}/2,\\
 t_{\rm pc}^{1,2} = &\; t+\mathrm{d}t/2,\\
 Z_{\rm p}^{1,2} = &\; \rho_{\rm Z}/\rho.
\end{split}\end{equation}
Here $\mathrm{d}t$ is the hydro time-step, and $\ub_{\rm rnd}$ is a random velocity (in opposite directions for $p^1$ and $p^2$ to conserve total momentum). 
This random velocity component is intended to reflect the unresolved motions of the cold clumps, which the stars originate from. 
Its absolute value is drawn from a Gaussian distribution with expectation value $0$ and variance $\sigma^2_{\ub_{\rm rnd}}$
\begin{equation}
 \sigma_{\ub_{\rm rnd}}^2=f_{\ub_{\rm rnd}}2K\left(1-\left(\frac{\ell_{\rm c}}{\Delta}\right)^{2\eta}\right).
\end{equation}
The fudge factor $f_{\ub_{\rm rnd}}=(\left<m_{\rm p}\right>m_{\rm p,th})/m_{\rm p}^2$ is designed to obtain $f_{\ub_{\rm rnd}}^{-2}\ub_{\rm rnd}$ as random velocity component at the end of the particle growth phase, when the final $m_{\rm pi}\approx(\left<m_{\rm p}\right>m_{\rm p,th})^{1/2}$ is reached. 
$\left<m_{\rm p}\right>$ is the mean mass of all stellar particles and $m_{\rm p,th}$ the upper threshold mass for particle growth.\\
A newly created stellar particle $p$ collects the stellar mass in $\rho_{\rm s,m}$ along its path [using a Nearest Grid Point algorithm (NGP)]. Its properties are updated using
\begin{equation}\begin{split}
 \acute{m}_{\rm p} = &\; m_{\rm p}+\rho_{\rm s,m}\Delta^{3},\\
 \acute{\ub}_{\rm p} = &\; \frac{\ub_{\rm p}m_{\rm p}+\ub\rho_{\rm s,m}\Delta^{3}}{\acute{m}_{\rm p}},\\
 \acute{m}_{\rm pi} = &\; m_{\rm pi}+\rho_{\rm s,m}\Delta^{3},\\
 \acute{t}_{\rm pc} = &\; \frac{t_{\rm pc}m_{\rm p}+(t+\mathrm{d}t/2)\rho_{\rm s,m}\Delta^{3}}{\acute{m}_{\rm p}},\\
 \acute{Z}_{\rm p} = &\; \frac{Z_{\rm p}m_{\rm p}+\frac{\rho_{\rm Z}}{\rho}\rho_{\rm s,m}\Delta^{3}}{\acute{m}_{\rm p}},\\
 \acute{\xb}_{\rm p} = &\; \frac{\xb_{\rm p}m_{\rm p}+\xb\rho_{\rm s,m}\Delta^{3}}{\acute{m}_{\rm p}}.
\end{split}\end{equation}
The final mass is reached, when either $m_{\rm p}>m_{\rm p,th}\simeq20\times m_{\rm p,min}$ or $(t-t_{\rm pc})>2\mbox{Myr}$.
\subsubsection{Feedback mechanism and hot phase}\label{sec:fb_hot_phase}
We consider the two physical stellar feedback processes, Lyman continuum heating and SNe explosions, and use the equations in BS12 to compute their contributions to the source terms for the update of the hydrodynamical state.\\
The stellar mass sink field $\rho_{\rm s,m}$ acts on the gas only via Lyman continuum heating $\Gamma^{\rm Lyc,m}$, where we assume zero age of the stellar population it represents. 
The amount of feedback (the mass of SNe ejecta $m_{\rm p,fb}$, the mass in the different chemical species $m_{\rm p,X,fb}$, and the heating rate due to Lyman continuum radiation $\Gamma^{\rm Lyc,p}$) during a hydro time-step $\mathrm{d}t$ is computed individually for every stellar particle according to its properties. 
To obtain $\dot{\rho}_{\rm s,FB}$, the feedback is mapped to the hydro-mesh using a cloud in cell (CIC) scheme and time-averaged over the time step $\mathrm{d}t$. The SNe ejecta mass is computed by
\begin{equation}
  \label{eq:sn_feedback}
 m_{\rm p,fb} = 
 m_{\rm pi}\int_{t-t_{\rm pc}}^{t-t_{\rm pc}+\mathrm{d}t}
 m_*\frac{\mathrm{d}N_{\rm *}}{\mathrm{d}m_{\rm *}}
 \frac{\mathrm{d}m_{\rm *}}{\mathrm{d}t'}\,\mathrm{d}t',
\end{equation}
where $m_{\rm *}=m_{\rm *}(Z_{\rm p},t')$ is the initial mass of an individual star that goes SN at an age of $t'$ \citep{Raiteri1996}. $m_{\rm p,fb}$ is removed from the particle mass: $\acute{m}_{\rm p}=m_{\rm p}-m_{\rm p,fb}$. 
The metal load of the ejecta is given by $m_{\rm p,Z,fb}=m_{\rm p,fb}(Z_{\rm p}+\zeta)$, where $\zeta$ is the fraction of metals produced in the dying stars, the other species scale linearly with metallicity. 
The total Lyc heating rate is $\Gamma^{\rm Lyc}=\Gamma^{\rm Lyc,m}+\Gamma^{\rm Lyc,p\rightarrow m}$. 
A fraction of energy load of the ejecta $e_{\rm SN}\left.\frac{\partial\rho_{\rm s}}{\partial t}\right|_{\rm FB}$ is deposited into $\rho E$ (via $\rho e$) and $\rho K$, $(1-\epsilon_{\rm SN})$ and $\epsilon_{\rm SN}$, respectively. 
However, this does not account for the kinetic energy of the ejecta that must also be transferred to the hydro-mesh along with their mass: 
\begin{equation}
 \label{eq:kinFB}
\begin{split}
\left.\frac{\partial (\rho_{\rm s} E_{\rm kin})}{\partial t}\right|_{\rm FB} =&\sum_{p}\frac{\ub_{\rm s,p}^2}{2}\left.\frac{\partial\rho_{\rm s}}{\partial t}\right|_{\rm FB,p}.
\end{split}
\end{equation}
Here we sum over all local contributions to kinetic energy from every individual particles '$p$'
The momentum of the ejecta transferred to bulk momentum of the gas changes the bulk kinetic energy of the gas by 
\begin{equation}
 \label{eq:ekin_cov}
\begin{split}
\left.\frac{\partial (\rho E_{\rm kin})}{\partial t}\right|_{\rm FB} =\frac{\ub}{2}\left(2\ub_{\rm s}-\ub\right)\left.\frac{\partial\rho_{\rm s}}{\partial t}\right|_{\rm FB}.
\end{split}
\end{equation}
To conserve energy, we add the difference to $\rho K$
\begin{equation}
 \label{eq:excess}
\begin{split}
\left.\frac{\partial (\rho K)}{\partial t}\right|_{\rm FB}^{\rm ex} =&\left.\frac{\partial (\rho E_{\rm kin})}{\partial t}\right|_{\rm FB} -\left.\frac{\partial (\rho E_{\rm kin})}{\partial t}\right|_{\rm FB}.
\end{split}
\end{equation}
The heating of the gas to very high temperatures by SNe is described by a hot phase density $\rho_{\rm h}$, obeying the conservation equation
\begin{equation}\label{eq:hdens}\begin{split}
 \frac{\partial\rho_{\rm h}}{\partial t} + \nabla \cdot \left(\rho_{\rm h} \ub\right) + &\min\left[\rho_{\rm h}\nabla \cdot \ub,0\right]\\
  &=\left.\frac{\partial\rho_{\rm s}}{\partial t}\right|_{\rm FB}-\rho_{\rm h}\exp\left(-t/\tau_{h}\right),
\end{split}\end{equation}
where $\min\left[\rho_{\rm h}\nabla \cdot \ub,0\right]$ describes the loss of thermal energy due to adiabatic expansion\footnote{Employing a ceiling $(\rho_{\rm h}\nabla \cdot \ub)_{\rm max}=0$ prevents producing hot phase when gas is compressed.}, and $\rho_{\rm h}\exp(-t/\tau_{h})$ the decay of $\rho_{\rm h}$ due to successive mixing of the hot gas into the ISM. 
The half-lifetime-scale $\tau_{h}/\log(2)$ is defined such that a SNe bubble shell at typical expansion velocity (roughly the speed of sound in the hot phase) travels roughly $~1\;\mbox{kpc}$ during that period, which leads to $\tau_{h}\approx1\;\mbox{Myr}$. 
$e_{\rm h}\approx0.1\times e_{\rm SN}$ is the constant specific thermal energy of the hot phase gas.\\
For consistency, the input parameters $\hat{\rho}_{\rm w}$ and $\hat{e}_{\rm w}$ to the derivation of the radiative cooling rate $\Lambda_{\rm w}^{\rm rad}$ (equation~\ref{eq:lambw}) are computed as follows:
\begin{equation}\begin{split}
 \hat{\rho}_{\rm w}=&\rho_{\rm w}-\rho_{\rm h},\\
 \hat{e}_{\rm w}=&\frac{\rho_{\rm w}e_{\rm w}-\rho_{\rm h}e_{\rm h}}{\hat{\rho}_{\rm w}}.
\end{split}\end{equation}
\section{Simulations}\label{sec:simulations}
\subsection{Initial conditions}\label{sec:initial_cond}
In the simulation domain with a volume of $0.5\;\mathrm{Mpc}^3$ we initialize an isothermal gaseous disc with an exponential surface density profile residing in a static dark matter halo using the potential-method of \citet{WangKless10}, which gives initial conditions similar to \citet{AgerLake09}. 
The choice of this setup is advantageous compared to a setup using a constant vertical scale height of the disc alike that by \citet{TaskTan09}, because it is adiabatically stable. 

In the absence of a stellar component, the exponential gaseous disc is defined by its mass $M_{\rm gas}=10^{10}\;\mathrm{M}_\odot$, its orientation of the disc angular momentum $\nb_{\rm gas}$ assuming a radial scale length $r_{\rm gas}=3.5\;\mathrm{kpc}$, an initially uniform metallicity $Z_{\rm gas}=0.1\times Z_\odot$ and a temperature $T_{\rm gas}=4\times10^4\;\mathrm{K}$ of the disc.

The dark matter is modelled by a static halo with a NFW-shaped density profile \citep{NFW}. It only contributes to the dynamics via its gravitational acceleration
\begin{equation}\label{eq:dm_accel}
 \gb_{\rm dm}=\frac{-GM_{\rm dm}\rb}{\log(1\!+\!c_{\rm dm})\!-\!\frac{c_{\rm dm}}{1+c_{\rm dm}}}\left(\frac{\log(r_{\rm s})}{r^3}\!-\!\frac{c_{\rm dm}}{r_{\rm dm}r^2r_{\rm s}}\right),
\end{equation}
at a given position with distance vector $\rb$ from the halo centre, its absolute value $r$ and the scaled dimensionless radius $r_{\rm s}=(1+rc_{\rm dm}/r_{\rm dm})$. The NFW~profile used is fully characterized by the halo mass $M_{\rm dm}=10^{12}\;\mathrm{M}_\odot$, the virial radius $r_{\rm dm}=213\;\mathrm{kpc}$, and a concentration parameter $c_{\rm dm}=12$.
\subsection{Individual runs}\label{sec:runs}
We performed eight isolated galaxy runs with model parameters listed in Table~\ref{tab:runs}. 
All runs were carried out with a root grid of $256^3$ cells. 
AMR levels with a factor of 2 spatial and temporal refinement were created using refinement criteria based on density; specifically, any cells with density above the minimum value of $0.01\ \mathrm{M_\odot/pc^3}$ were tagged for refinement up unto a specified maximum number of levels. 
In runs ref, nE, nB, nEnB, sSF, and sSF2 a spatial resolution of $\sim30\ \mathrm{pc}$ was obtained using six levels of refinement. 
The effects of the feedback implementation are explored with the runs nE, nB, and nEnB in comparison with run ref, that features MIST with the reference parameters as given in Table~\ref{tab:coeffs_n_params}.\\
The runs sSF and sSF2 run feature a simplified model for the ISM without phase separation and $\epsilon_{\rm SN}=0$.
A threshold controlled star formation recipe is applied here, according to which stars are allowed to form at a free fall time efficiency of $\varepsilon_{\rm sSF}=0.01$, if the local density exceeds $\rho_{\rm sSF,min}=50m_{\rm H}\ cm^{-3}$ and the local temperature is lower than $T_{\rm sSF,max}=1.5\times10^4\ \mathrm{K}$. 
While in sSF2 the whole SGS-turbulence model is switched off, the model is still active in sSF, but star formation is decoupled from $K$ and the SGS-energy production terms related to MIST are switched off in this case (i.e. $\Pi_{\rm TI}\equiv0$ and $\Pi_{\rm SN}\equiv0$). Stellar particles are not created as pairs (see Section~\ref{sec:spart_creation}) in sSF and sSF2, since $\ell_{\rm c}$ is not defined in both cases.\\
In addition two runs with fewer refinement levels were performed in order to investigate the effects of numerical resolution on the results of our simulations. 
The runs lres5 and lres4 feature effective resolutions of $\sim60\ \mathrm{pc}$ and $\sim120\ \mathrm{pc}$ using five and four levels of refinement, respectively.\\
Up to densities around $0.3\ \mathrm{M_\odot\ pc^{-3}}$ the resolution requirement by \citet{Truelove1997} is easily satisfied in all runs with MIST. This limit can be shifted towards much higher densities, if we consider the effective pressure instead of thermal pressure only. 
The Jeans length of the few densest cells may temporarily drop below $4\Delta$ though, but never below the size of a cell $\Delta$, before the dense region is disrupted again by feedback. The latter statements are not true in case of sSF and sSF2, in which the Jeans length criterion is always violated in the dense clusters.\\
The combined usage of LES and MIST in a simulation increases the amount of computational resources required by less than 10 per cent compared to runs without, and significantly less than an additional level of refinement (more than 100 per cent).\\
The isolated galaxies were evolved for at least one orbital time at 10~kpc radius from the centre ($\sim400$~Myr).\\
\begin{table}
\caption{Simulation runs.}
\label{tab:runs}
\begin{tabular}{lllllll}
\hline
ID&$\Delta$&LES&ISM&$\epsilon_{\rm SN}$&$\rho_{\rm h}$&Stop time\\
\hline
\hline
ref&30 pc&Yes&MIST&0.085&Yes&1.0 Gyr\\
nE&30 pc&Yes&MIST&0.0&Yes&1.0 Gyr\\
nB&30 pc&Yes&MIST&0.085&No&0.5 Gyr\\
nBnE&30 pc&Yes&MIST&0.0&No&0.4 Gyr\\
sSF&30 pc&Yes&Simple&0.0&Yes&0.8 Gyr\\
sSF2&30 pc&No&Simple&0.0&Yes&0.4 Gyr\\
lres5&60 pc&Yes&MIST&0.085&Yes&1.0 Gyr\\
lres4&120 pc&Yes&MIST&0.085&Yes&1.0 Gyr\\
\hline
\hline
\end{tabular}
\end{table}

\section{Results}\label{sec:result}
\subsection{Disk evolution in the ref run}\label{sec:refrun}
\begin{figure*}\begin{minipage}{177mm}
\centering
  \resizebox{0.95\hsize}{!}{\includegraphics{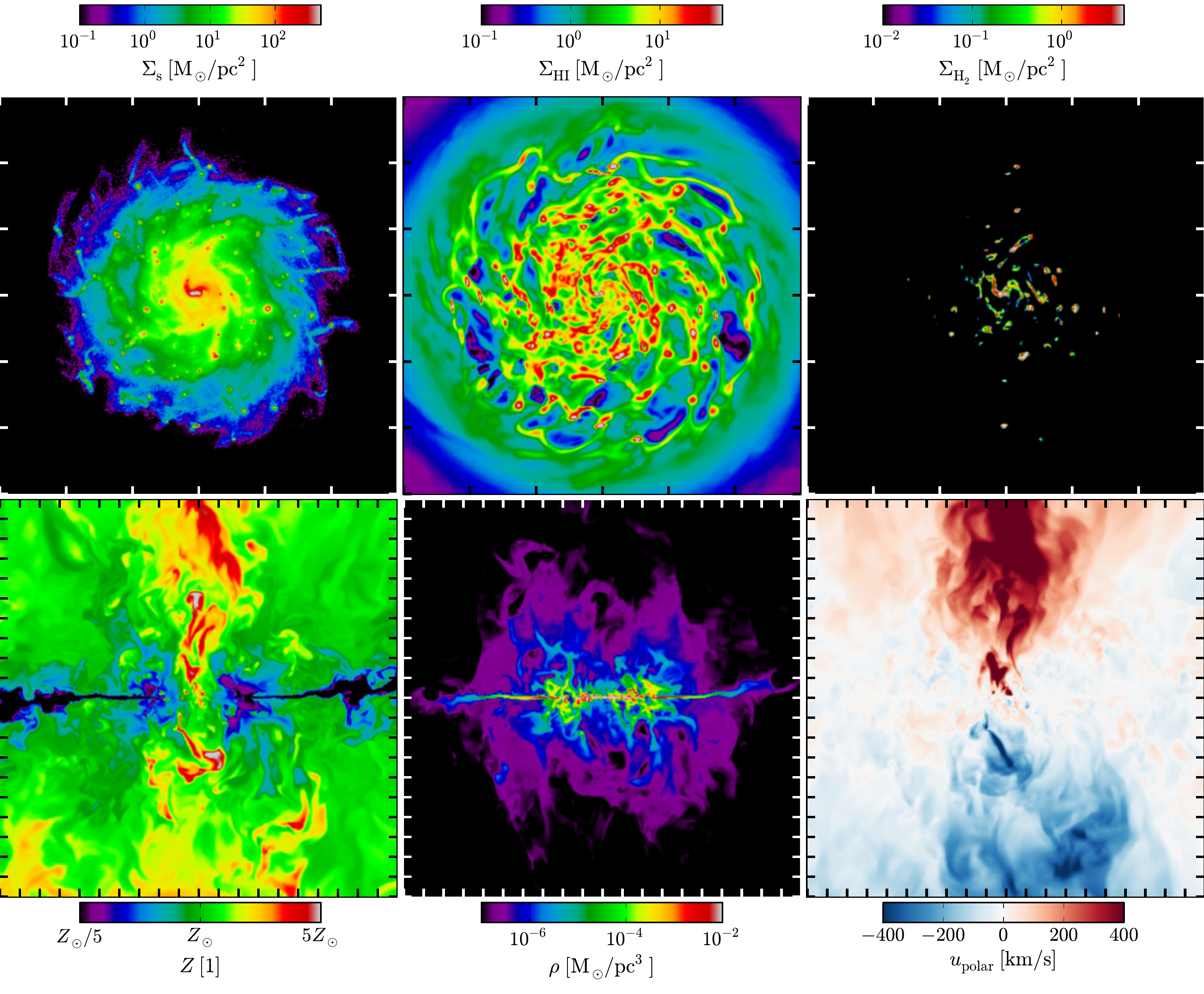}}
\caption{The top row of panels shows the following projected quantities in the central ($30\mathrm{kpc}\ \times\ 30\mathrm{kpc}$) region of the galactic disc of the ref run after 1~Gyr (left to right): stellar column density $\Sigma_{\rm s}$, $HI$ column density $\Sigma_{\rm HI}$, star formation column density $\dot{\Sigma}_{\rm SF}$. The bottom row shows polar ($100\mathrm{kpc}\ \times\ 100\mathrm{kpc}$) slices of the central region of the galactic disc of the ref run at 1~Gyr (left to right): metallicity $Z$, total gas density $\rho$, velocity perpendicular to the disc plane $v_{\rm polar}$. The tick marks on the plot edges have a spacing of 5~kpc.}
\label{fig:multi_colored}\end{minipage}
\end{figure*}
Initially the disc is adiabatically stable, but as the gas is allowed to cool by radiation, it loses its thermal support in height and collapses into a thin cold disc. 
The disc becomes Toomre-unstable and fragments into clumps. In those clumps the gas eventually becomes dense enough to become molecular and consequently begins to form stars. 
The SNe feedback of the newly formed stars then eventually disperses the clumps. 
A fraction of the stars formed in some of those clumps may form a stellar cluster that survives over a much longer time than the lifetime of about 20~Myr of the gas clump from which it originated. 
The stellar clusters tend to move towards the centre of the disc as a result of dynamical friction, where they eventually merge into a central agglomeration of stars, if they are not disrupted before by the tidal forces in the disc. 
However, they do not build up a bulge as their velocity dispersion is too small. 
The majority of the stars form a rather smooth disc with a scale height of a few hundred parsec. 
The structure of the stellar disc is shown in the top left panel of Fig.~\ref{fig:multi_colored}. 
The stars stripped in the potential of the disc form tidal tails around their birth cluster.\\

The SNe also carve holes into the disc and launch a wind leaving the disc. 
The interplay between cooling, gravity, and SNe shapes the gaseous component into a fluffy disc, with holes and knots, as demonstrated in the top and bottom panels in the middle of Fig.~\ref{fig:multi_colored}. 
The wind mostly not only consists of metal enriched hot gas from SNe, but also carries a fraction of the original cold clump with it (see bottom left panel of Fig.~\ref{fig:multi_colored}). 
The wind originates from the cavities of the disc caused by the SNe at speeds ranging from 300 to $1000\ \mathrm{km\ s}^{-1}$, and pushes a shell of cold or warm gas outward. 
The ejected gas is either mixed into the wind, or falls back into the disc. 
Far away from its origin, the wind from all sources merges into a hot, but dilute, sub-sonically turbulent, and metal-rich ($\sim Z_\odot$) bubble that continues to expand. 
The bottom panels of Fig.~\ref{fig:multi_colored} give an impression of the metallicity, mass, and velocity structures in the winds above the disc.\\

\begin{figure}
\centering
  \resizebox{1.0\hsize}{!}{\includegraphics{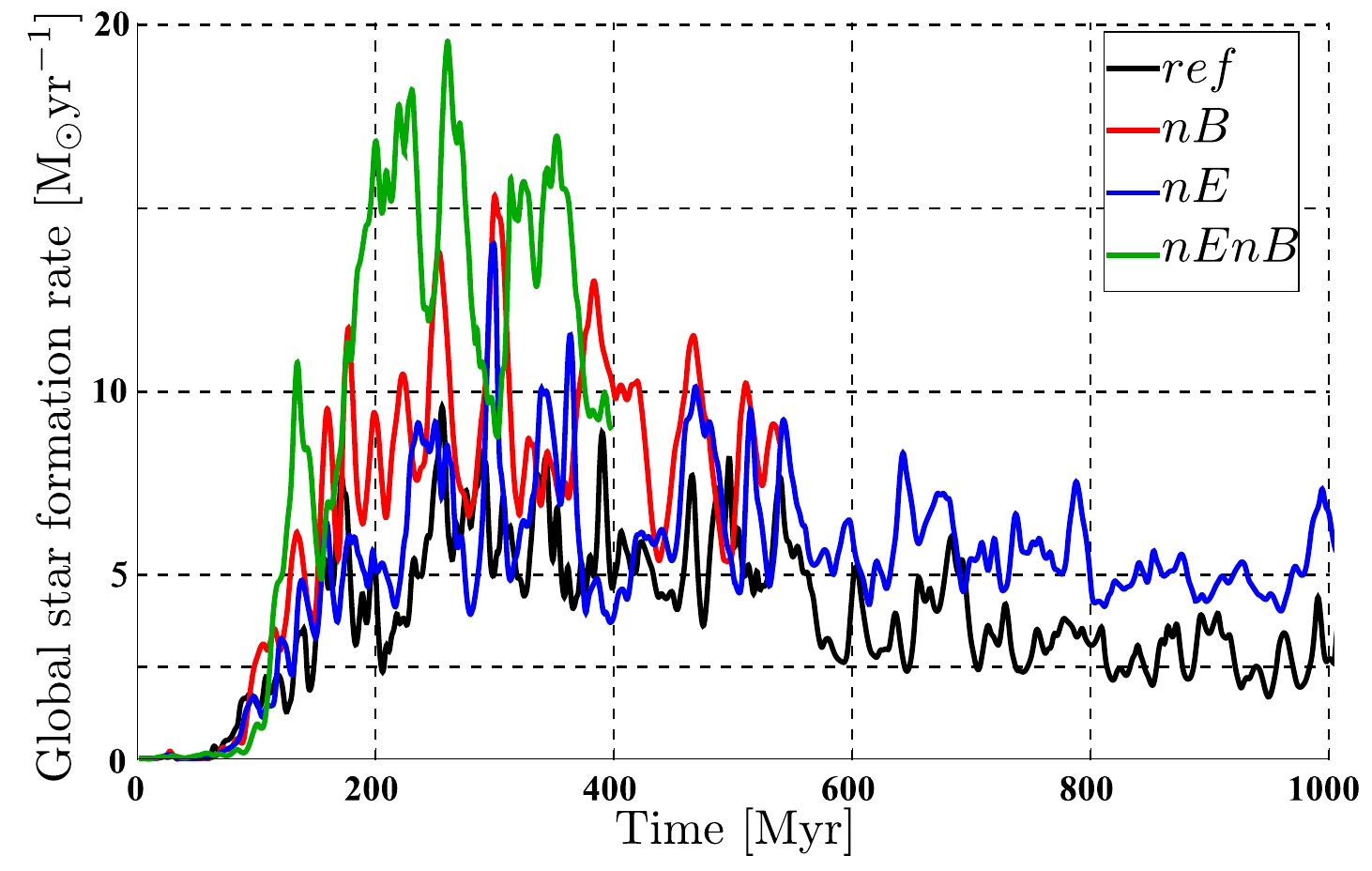}}
\caption{Global star formation rate $\dot{M}_{\rm SF}$ over simulation time for the different runs ref, nB, nE, and nEnB in black, red, blue, and green, respectively.}
\label{fig:globalSFR}
\end{figure}

As seen in the top right panel of Fig.~\ref{fig:multi_colored}, active star formation occurs only in a few compact regions away from the centre. 
Because of the local metal enrichment due to SNe ejecta, the threshold density for star formation drops. 
This causes clumps to form stars earlier during their collapse, when the density is still relatively low, which lowers their star formation rate and makes them more prone to dispersal by SNe. 
The residual stellar clusters are fewer, lighter, and more easily disrupted in the galaxy's potential.\\
The global star formation rate in the ref run is plotted as black line in Fig.~\ref{fig:globalSFR}. 
Initially the amount of star formation increases quickly, as the region of star formation grows. 
It reaches its peak around 300~Myr after start of simulation, and then gradually declines due the consumption and the metal enrichment of the gas reservoir in the inner disc.
At this stage approximately 30 per cent of the initial gas mass has been converted into stars.
\subsubsection{Differences in the nE run}
\begin{figure}
\centering
  \resizebox{1.0\hsize}{!}{\includegraphics{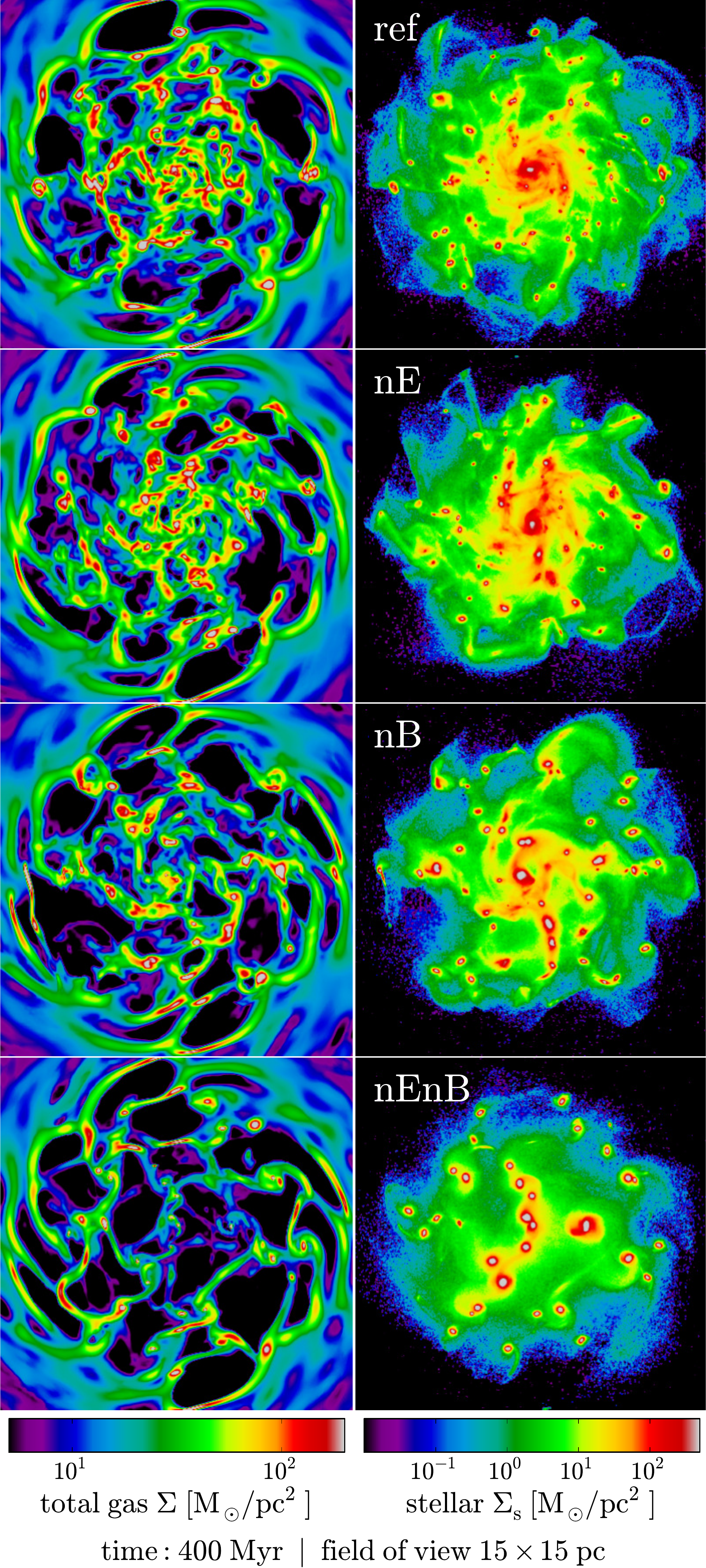}}
\caption{Comparison of total gas surface density $\Sigma$ in the left column and the stellar surface density $\Sigma_{\rm s}$ in the right column between the different runs (from top to bottom: ref, nE, nB, nEnB) 400 Myr after start of simulation.}
\label{fig:runs_synopsis}
\end{figure}
In the nE run we set $\epsilon_{\rm SN}=0$. 
Thus SNe feedback does not directly increase the unresolved turbulent energy $K$ (see equation~\ref{eq:turb}). This has basically two effects on the overall evolution. 
On the one hand, star formation in a clump is active over a longer period of time because of the higher star formation efficiency in moderately turbulent clumps ($\mathcal{M}_{\rm c}\sim 10$), causing more stars to form before a clump of gas is dispersed. 
This leads to a slightly higher global star formation rate than in the ref run (see the blue line in Fig.~\ref{fig:globalSFR}).
On the other hand, stars forming in a clump after the first stars have already produced SNe have a significantly reduced velocity dispersion compared to their analogues from the ref run. 
As a consequence, those stars are more likely to stay near the clump of their origin. In this case, stellar clusters tend to be more massive and more strongly gravitationally bound, and hence, their tidal tails are less prominent. 
Clusters are initially more abundant and have a longer lifetime. Eventually they merge into a few very massive clusters, which accrete gas, and host intermittent star formation, as the gas is driven apart due to feedback. 
They remain gravitationally bound and stable, as their stellar mass is sufficiently large and concentrated.\\
As a consequence, the resulting stellar and gaseous discs are more clumpy than in the ref run. 
This can be seen by comparing the projections of total gaseous density and stellar density in Fig.~\ref{fig:runs_synopsis}, where both stellar and gaseous clumps are fewer in numbers of appearance but more massive with higher central densities. 
During the simulation approximately 40 per cent of the initial gas mass was consumed by star formation after 1~Gyr.
\subsubsection{Differences in the nB run}
In the nB run we turned off the treatment of the hot SN ejecta phase, i.e. the hot SN gas is directly mixed into the warm phase. 
This allows the gas to radiate away the feedback energy more quickly. 
Thus star formation in a clump goes on for a longer time, consuming a larger fraction of the gaseous mass. 
Once almost all gas is depleted the feedback takes the lead, and drives stronger and faster winds than in the ref run. 
Like in the nE run, the resulting stellar clusters are more stable and massive, and subsequently merge to form very massive clusters, that cannot be disrupted in the tidal field of the disc. 
Those massive clusters are hosting continuous star formation. 
They keep accreting gas from the disc and converting it into stars, thereby sustaining their gaseous mass at the same level for a long time.\\
As in the nE run, the resulting gaseous and stellar discs are clumpy, but the stellar clusters are much more massive, and the gas is depleted on much shorter time-scales (see Fig.~\ref{fig:runs_synopsis}). 
The stellar density in the centre of the clusters reaches values much greater than $10^3\ \mathrm{M_\odot\ pc^{-3}}$. 
This is why we stopped this simulation after 0.5~Gyr. During this period of time about 30 per cent of gaseous mass was turned into stars.
\subsubsection{Differences in the nEnB run}
Setting $\epsilon_{\rm SN}=0$ and turning off the treatment of the hot SN ejecta phase combines the effects described above. 
This leads to an even more violent evolution with a few very massive stellar clusters in the inner regions, while the gas is depleted quickly (see Fig.~\ref{fig:runs_synopsis}). 
We stopped this run even earlier than the nB run. The amount of gas consumed by star formation was about 30 per cent in roughly 400~Myr.
\subsection{Star formation}
\subsubsection{Time scales}\label{sec:tdep}
\begin{figure}
\centering
  \resizebox{0.625\hsize}{!}{\includegraphics{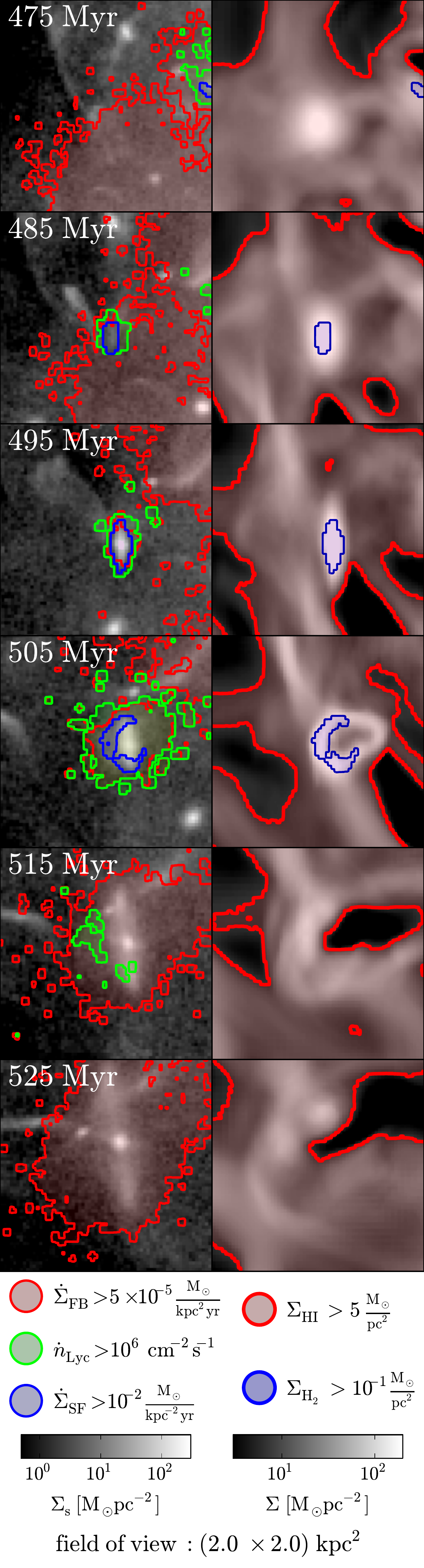}}
\caption{Evolution of a gas clump from shortly before star formation begins until dispersal. In the left panels the stellar surface density $\Sigma_{\rm s}$ in grey-scales is over-plotted with contours of the surface density of the SNe feedback $\dot{\Sigma}_{\rm FB}$ in red (indicating stellar populations at ages between $\sim$4 and $\sim$40~Myr), the surface density of Lyman continuum emission $\dot{n}_{\rm Lyc}$ in green (indicating stellar populations younger than $\sim$2~Myr), and the star formation surface density $\dot{\Sigma}_{\rm SF}$ in blue. 
Right panels show the total gas surface density $\Sigma$ in grey-scales over-plotted with contours of surface density of cold gas $\Sigma_{\rm HI}$ in red, and the surface density of dense molecular gas $\Sigma_{\rm H_2}$ in blue.}
\label{fig:clump_time_series}
\end{figure}
\begin{figure}
\centering
  \resizebox{1.0\hsize}{!}{\includegraphics{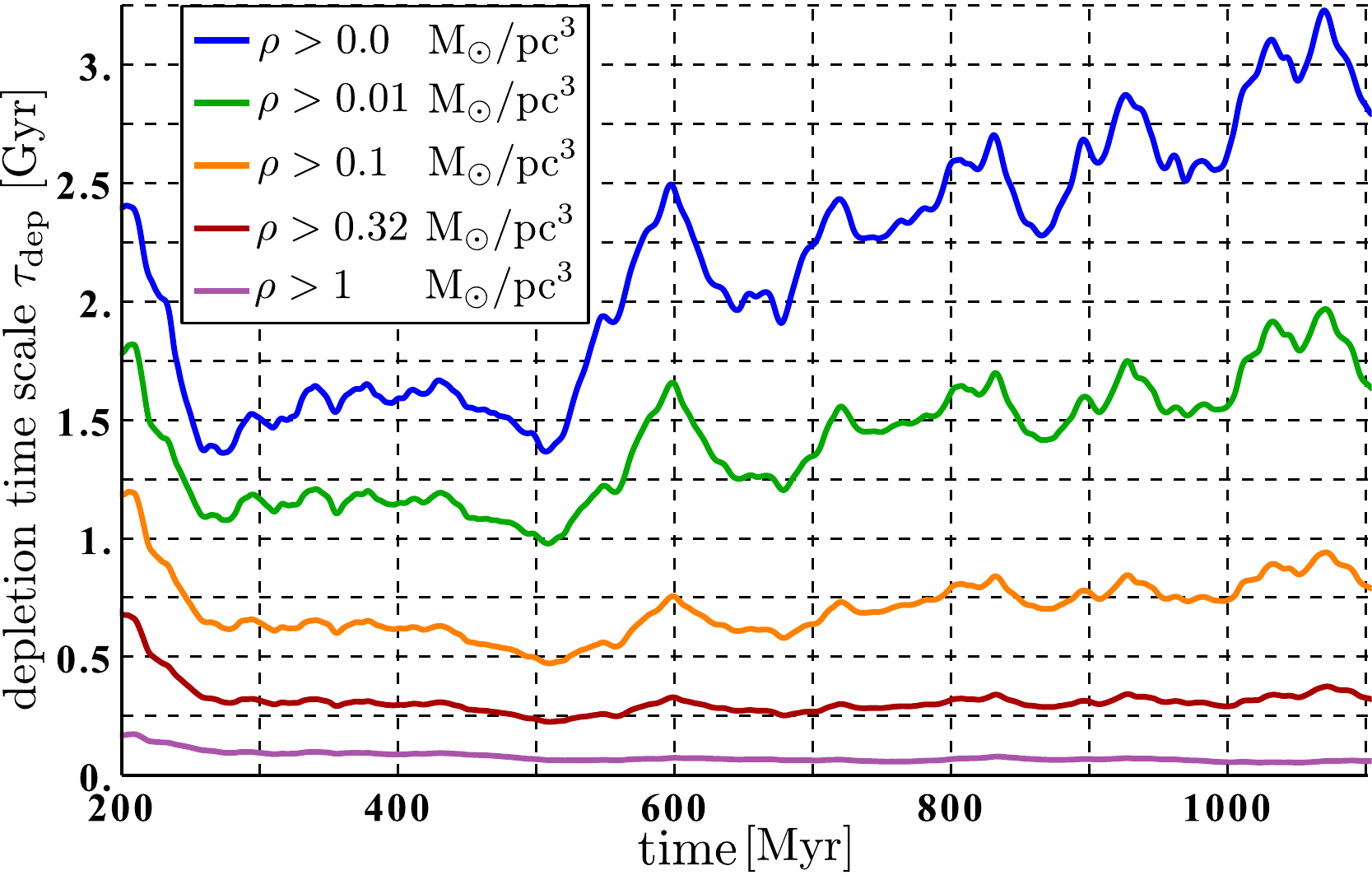}}
\caption{Global gas depletion time-scale $\tau_{\rm dep}$ versus time in the ref run. 
We plot moving averages of $\tau_{\rm dep}$ to eliminate short period variations caused by single star-forming regions. 
$\tau_{\rm dep}$ was derived with respect to gaseous mass in regions with $\rho>\rho_{\rm thr}$. 
Blue, green, orange, red, and purple lines show  $\tau_{\rm dep}$ using threshold densities $\rho_{\rm thr}=\{0.0,0.01,0.1,0.32,1\}\ \mathrm{M_\odot\ pc^{-3}}$, or in terms of number density $n_{\rm thr}\simeq\{0.0,0.33,3.3,10,33\}\ \mathrm{cm^{-3}}$.}
\label{fig:globalTdep}
\end{figure}

The $\sim$20~Myr time-scale related to single star-forming regions reflects the period of time needed to produce enough stars, such that their combined feedback is able to quench star formation, and to evaporate the most dense, central region of the star-forming cloud. 
If the environment is still dense enough, star formation may continue in an compressed layer around the expanding bubble. Depending on the mass of the clump and its surroundings, star formation can continue in this mode. 
A time series depicting the evolution of one of those clumps is shown in Fig.~\ref{fig:clump_time_series}. 
The impact of the most massive clumps is seen in the quick variations of the global star formation rate $\dot{M}_{\rm SF}$ on time-scales around 10 to 40~Myr (see Fig.~\ref{fig:globalTdep}). 
Metal-enriched clouds tend to have a shorter lifetime compared to metal-poor clouds, as star formation can start at lower densities, because shielding from radiation is enhanced due to higher dust abundances.\\
By assuming different threshold densities $\rho_{\rm thr,i}=\left\{0.0,0.01,0.1,0.32,1.0\right\}\ \mathrm{M_\odot\ pc^{-3}}$ to compute the reservoir of available gas, we derive global gas depletion time-scales 
\begin{equation}
 \tau_{\rm dep,i}(t)=\frac{\sum_{\rho(t)>\rho_{\rm thr,i}}\rho\Delta^3}{\sum\dot{\rho}_{\rm SF}\Delta^3}=\frac{M_{\rm gas,i}}{\dot{M}_{\rm SF}}.
\end{equation}
To filter out quick variations, we apply a moving average 
\begin{equation}
 \overline{\tau}_{\rm dep,i}(t)=\frac{1}{t_{\rm av}}\int_{t-t_{\rm av}/2}^{t+t_{\rm av}/2}\tau_{\rm dep,i}(\grave{t})\mathrm{d}\grave{t}
\end{equation}
with $t_{\rm av}=50\ \mathrm{Myr}$.
Plots of the resulting depletion times $\tau_{\rm dep,i}$ are shown in Fig.~\ref{fig:globalTdep} for different choices of $\rho_{\rm thr,i}$. 
The typical time-scales range from less than 100~Myr for the highest threshold of 1 $\mathrm{M_\odot\ pc^{-3}}$ to a few Gyr for the total gas content of the simulation domain. 
Gas above densities $\rho_{\rm thr}=1\ \mathrm{M_\odot\ pc^{-3}}$ is most likely actively star-forming. 
Contrary to the other cases, $\tau_{\rm dep}$ for $\rho_{\rm thr}=1\ \mathrm{M_\odot\ pc^{-3}}$ slightly shrinks with time. 
Because of the increasing metallicity the threshold density needed for star formation drops gradually. 
Star formation and subsequent feedback then prevent gas from becoming as dense as in metal-poorer environments. 
$\tau_{\rm dep}$ for $\rho_{\rm thr}=0.32\ \mathrm{M_\odot\ pc^{-3}}$ remains almost constant at 0.25~Gyr after the initial transient phase. 
For lower $\rho_{\rm thr}$, $\tau_{\rm dep}$ grows with time, and the growth rate tends to increase with decreasing $\rho_{\rm thr}$. 
The growth of $\tau_{\rm dep}$ if $\rho_{\rm thr}\leq0.32\ \mathrm{M_\odot\ pc^{-3}}$ is also related to the increase of metallicity. 
In metal-rich environments gas forms stars already at lower densities, implying greater $\tau_{\rm ff,c}$, but the efficiency $\epsilon_{\rm PN}$ (see equation~\ref{eq:sfrMIST}) remains about the same. 
For this reason $\dot{M}_{\rm SF}$ is effectively lowered faster than gas supply $M_{\rm gas}$ shrinks. 
We expect that all $\tau_{\rm dep}$ will saturate, as the impact of increasing metallicity is further reduced in already enriched gas (see BS12). 
However, the magnitude of the depletion time-scales is well within range of observationally inferred ones \citep{Daddi2010, Genzel2010}. 
\subsubsection{Local efficiency}\label{sec:loceff}
\begin{figure*}
\begin{minipage}{177mm}
\centering
  \resizebox{0.95\hsize}{!}{\includegraphics{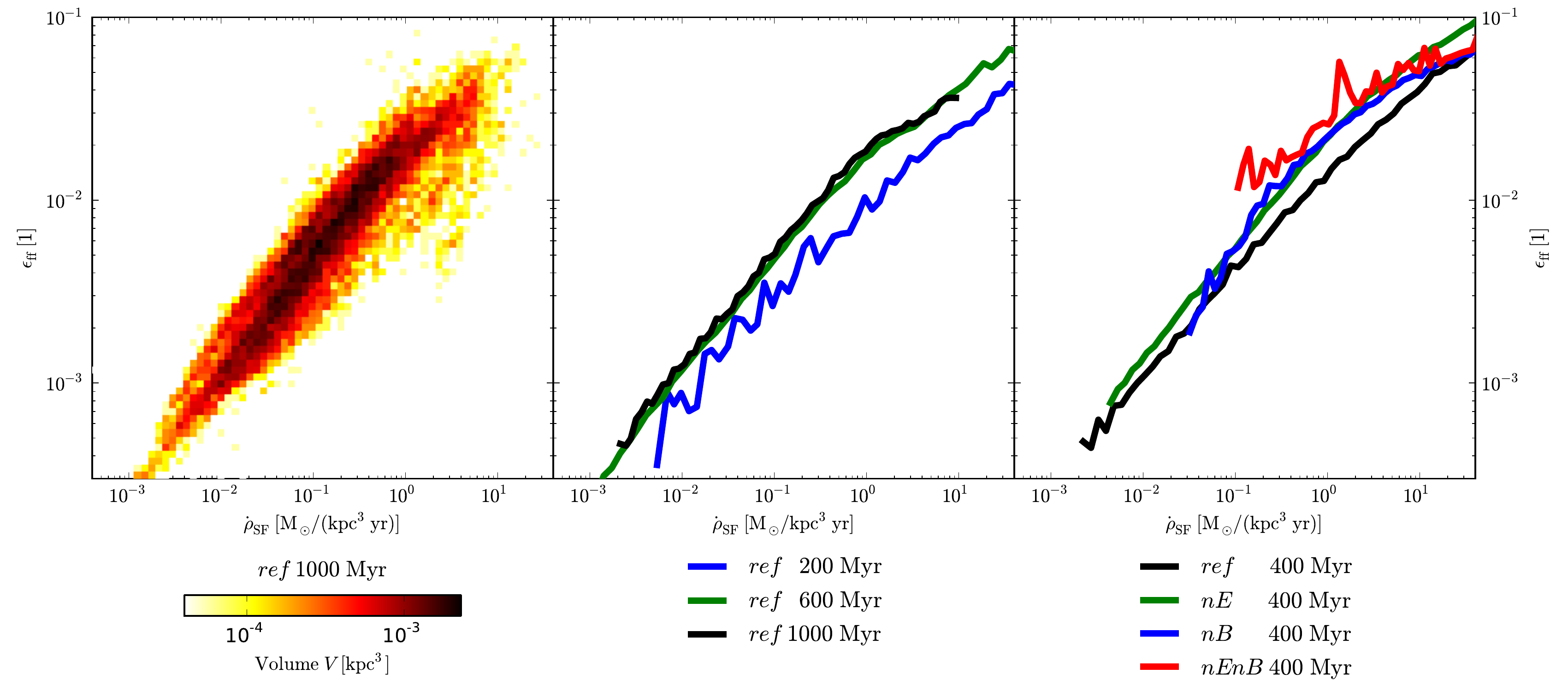}}
\caption{Star formation efficiency $\epsilon_{\rm ff}$ over star formation density $\dot{\rho}_{\rm SF}$. The left panel shows a two dimensional area weighted histogram of the ref-data after $1\ \mathrm{Gyr}$. The middle panel shows a time series (blue: 200~Myr, green: 600~Myr, black: 1~Gyr) of the mean $\epsilon_{\rm ff}$ corresponding to a specific value of $\dot{\rho}_{\rm SF}$, while on the right panel a comparison of the latter between the individual runs (black: ref, green: nE, blue: nB, red: nEnB) at 400~Myr is plotted.}
\label{fig:cPPr_SF_Eff}\end{minipage}
\end{figure*}
The star formation efficiency is defined by
\begin{equation}
 \dot{\rho}_{\rm SF} = \frac{\epsilon_{\rm ff}\rho}{\tau_{\rm ff}},
\end{equation}
where a constant value for $\epsilon_{\rm ff}\simeq0.01$ is commonly assumed. 
In our simulations $\epsilon_{\rm ff}$ is computed dynamically from the local turbulent hydrodynamical state. 
While the star formation efficiency in the cold phase, $\epsilon_{\rm PN}$ (see equation~(\ref{eq:sfrMIST})), is almost constant -- around $\epsilon_{\rm PN}\simeq0.1$ -- for all star-forming regions, the shielded molecular content $f_{\rm H_2}\rho_{\rm c}$ and $\rho_{\rm c,pa}$ vary significantly, such that $\epsilon_{\rm ff}$ is boosted in high density regions. 
As a consequence, a large contribution to the global star formation rate comes from just a few temporarily very active spots in the disc, which explains the relatively large variations on short time-scales. 
As shown in Fig.\ref{fig:cPPr_SF_Eff}, we find a good correlation $\epsilon_{\rm ff}\propto\sqrt{\dot{\rho}_{\rm SF}}$, which holds for all times in the ref run and the nE run. 
In the other runs without treatment of a hot phase, the power-law slope flattens towards large $\dot{\rho}_{\rm SF}$, which is clearly a sign that the effectiveness of thermal feedback plays a major role. 
The typical value of $\epsilon_{\rm ff}\simeq0.01$ is, on average, reproduced in runs with hot phase treatment, while those without produce stars at significantly higher average efficiency.
\subsubsection{$\mathrm{H_2}$ and $\mathrm{HI}$}\label{sec:H2HI}
\begin{figure*}
\begin{minipage}{177mm}
\centering
  \resizebox{0.95\hsize}{!}{\includegraphics{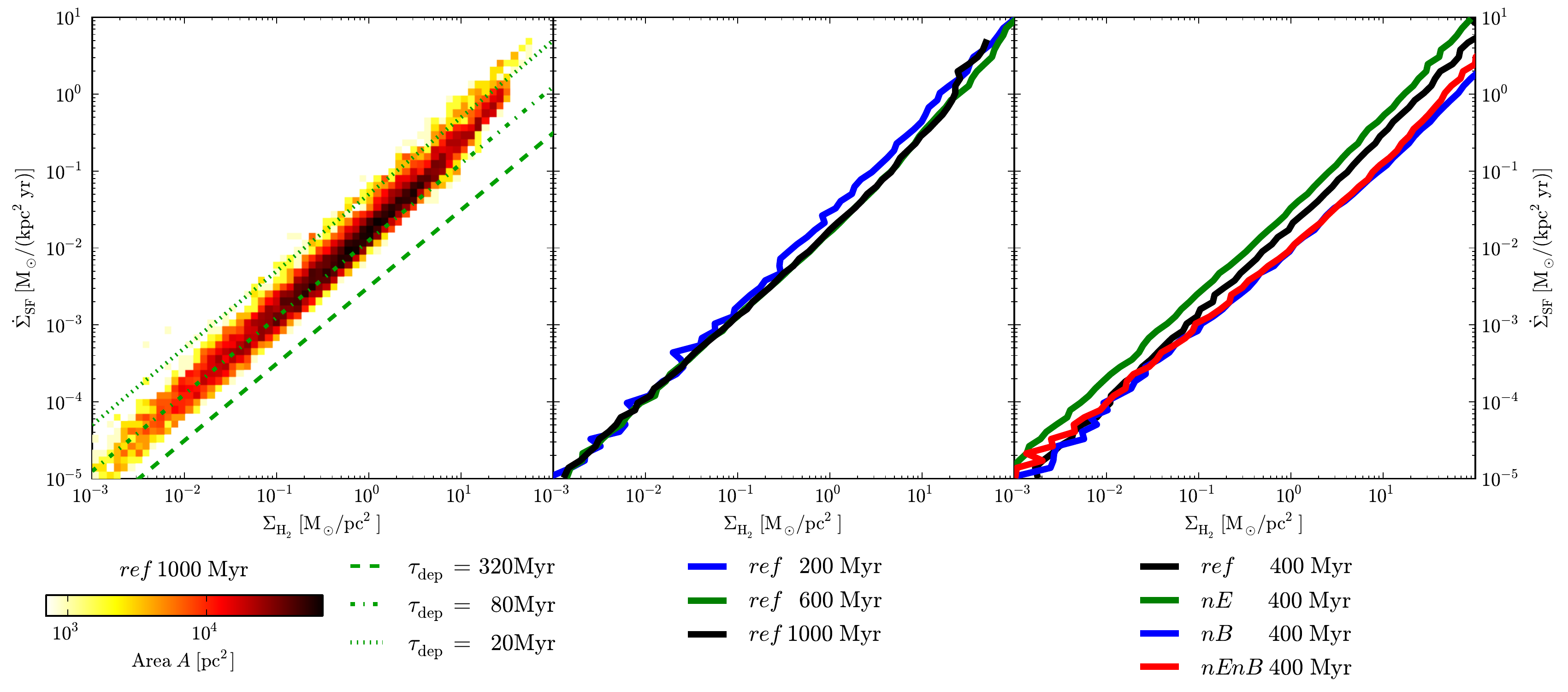}}
\caption{Star formation column density $\dot{\Sigma}_{\rm SF}$ over $\mathrm{H_2}$ column density $\Sigma_{\rm H_2}$. Plots and colours are arranged as in Fig.~\ref{fig:cPPr_SF_Eff}. Additionally, in the left panel the dotted, dash-dotted, and dashed green lines indicate constant depletion time-scales of 20, 80, and 320~Myr, respectively.}
\label{fig:comp_SPSPr_H2SF}\end{minipage}
\end{figure*}
We find a very tight correlation between the star formation column density $\dot{\Sigma}_{\rm SF}$ and the molecular column density $\Sigma_{\rm H_2}$, as shown on the left panel of Fig.~\ref{fig:comp_SPSPr_H2SF}. 
This correlation implies a robust power-law relation between $\dot{\Sigma}_{\rm SF}$ and $\Sigma_{\rm H_2}$ with an exponent $\alpha_{\rm H_2,SF}=1.05\pm0.06$ slightly above one. 
This corresponds to an almost constant depletion time of the molecular gas in star-forming regions around 80~Myr. 
This correlation holds for all runs independent of the simulation time, as shown in the middle and right panels of Fig.~\ref{fig:comp_SPSPr_H2SF}.\\

The existence of the correlation itself is not surprising, as it is assumed in the MIST model equations that $\dot{\rho}_{\rm SF}$ is proportional to $f_{\rm H_2}$ (see equation~\ref{eq:sfrMIST}). $\alpha_{\rm H_2,SF}\simeq 1$, however, is not imposed. 
Assuming a constant star formation efficiency $\epsilon_{\rm PN}$, one would expect $\alpha_{\rm H_2,SF}\simeq 1.5$, but in our model $\epsilon_{\rm PN}$ and $f_{\rm H_2}$ depend implicitly and non-linearly on $\rho_{\rm c,pa}$, $\mathcal{M}_{\rm c}$, and $l_{\rm c}$, which are in turn nonlinear functions of $\rho$, $\rho_{\rm c}$, $\rho_{\rm w}$, $K$, and $e_{\rm w}$. 
BS12 already showed that a correlation with $\alpha_{\rm H_2,SF}\simeq 1$ can be found in the equilibrium solutions of the MIST model equations. 
This suggests that the interplay of the implicit dependencies effectively results in an almost linear relation between estimated $\mathrm{H_2}$-mass available to star formation and the star formation rate, which is an important feature of the self-regulating mechanisms implemented in MIST. 
The depletion time-scale of molecular gas from the equilibrium solutions (BS12) is about 1-2~Gyr, in agreement with that found by e.g. \citet{Bigiel2011} from CO-observations. 
We derive an average depletion time of $\sim80$~Myr in our simulations, or shorter ($\sim40$~Myr) in regions of higher molecular density. 
This is considerably, by a factor of around 25, shorter (see Fig.~\ref{fig:comp_SPSPr_H2SF}). 
But this time-scale matches the depletion time of molecular gas within star-forming molecular clouds, found by \citet{Murray2011}\footnote{\citet{Murray2011} find a GMC star-forming efficiency $\epsilon_{\rm GMC}\simeq0.12$ and related free fall times $\tau_{\rm ff,GMC}\simeq10\ \mathrm{Myr}$. 
A combination of both yields a depletion time-scale of $\tau_{\rm dep,GMC}\simeq80\ \mathrm{Myr}$.}, and that for dense molecular gas inferred from HCN-observations by \citet{GaoSol2004}\footnote{\citet{GaoSol2004} find a linear correlation between the galactic star formation rate $\dot{\rho}_{\rm SF,gal}$ and the amount $M_{\rm dense}$ of what they call 'dense gas' in a galaxy: $\dot{\rho}_{\rm SF,gal}=1.8M_{\rm dense}/10^8\mathrm{yr}^{-1}$. 
This corresponds to a dense gas depletion time of $\sim60\ \mathrm{Myr}$.}, and the total gas depletion time-scale $\tau_{\rm dep}$ for $\rho_{\rm thr}=1\ \mathrm{M_\odot\ pc^{-3}}$. 
This discrepancy between the depletion time-scales of molecular gas is caused by the fact that we estimate the amount of gas involved in the processes of active star formation by calculating the molecular fraction of gas in the centres of cold cloud complexes in equilibrium with an external radiation field. 
However, there are also environments containing significant amounts of $\mathrm{H_2}$ that are not in equilibrium with radiation or only partially shielded. 
Molecular material that is transported from shielded to not shielded regions by turbulent motions or SNe blast waves without being instantaneously dissociated is not tracked in our model for example. 
This material, as well as gas that is partially molecular but not dense enough to form stars, contributes to observed molecular hydrogen column densities, but is not present in our simulations. 
Our data on the $\mathrm{H_2}$-content of the ISM do reflect the amount of dense molecular gas that is actively forming stars, but do not reproduce the full amount of molecular gas seen in CO-observations. 
Coarse graining our $\mathrm{H_2}$-data changes neither the inferred depletion time-scales nor the slope $\alpha_{\rm H_2,SF}$ significantly, which is consistent with \citet{GaoSol2004}, who used galactic quantities.\\
\begin{figure*}\begin{minipage}{177mm}
\centering
  \resizebox{0.95\hsize}{!}{\includegraphics{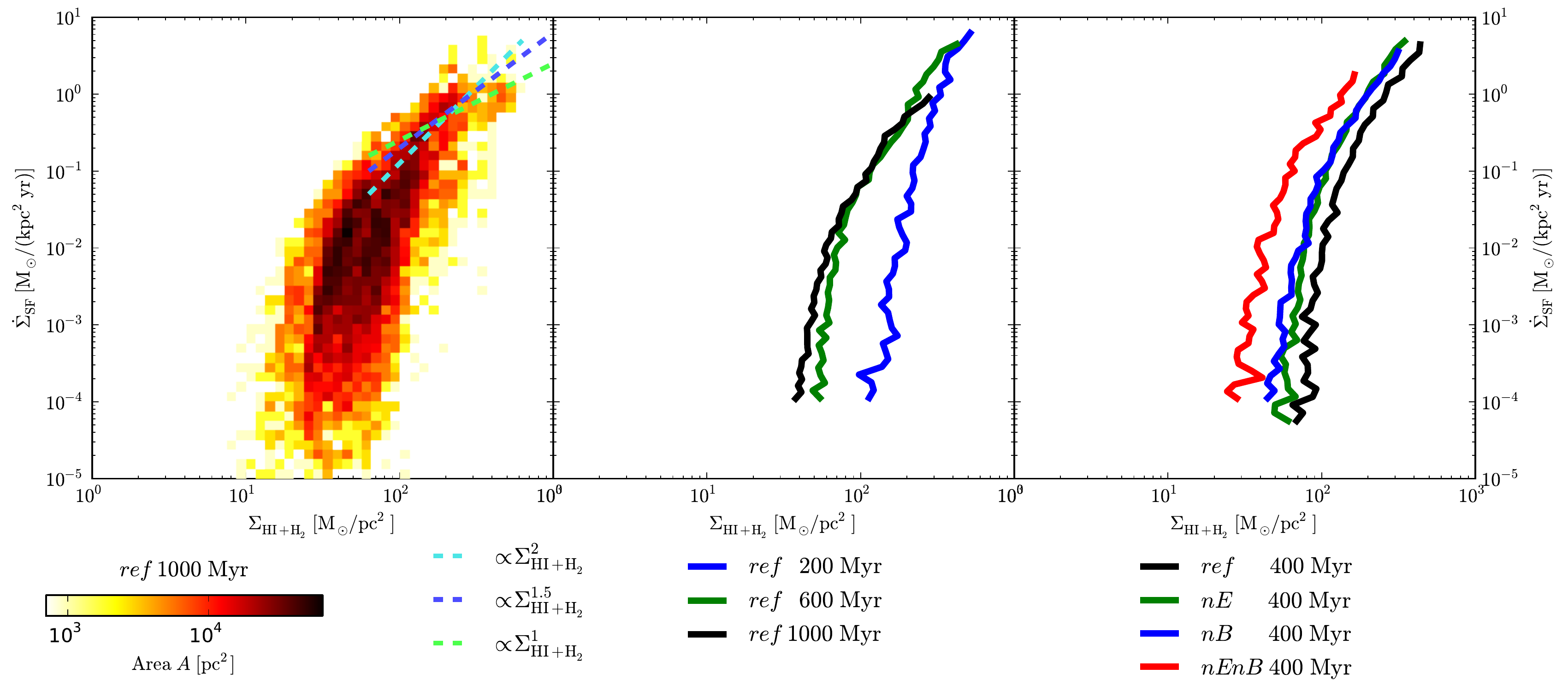}}
\caption{
Star formation column density $\dot{\Sigma}_{\rm SF}$ over $\mathrm{HI+H_2}$ column density $\Sigma_{\rm HI+H_2}$. 
Plots and colours are arranged as in Fig.~\ref{fig:cPPr_SF_Eff}. 
Fits $\dot{\Sigma}_{\rm SF}\propto\Sigma_{\rm HI+H_2}^{2}$, $\dot{\Sigma}_{\rm SF}\propto\Sigma_{\rm HI+H_2}^{1.5}$, and $\dot{\Sigma}_{\rm SF}\propto\Sigma_{\rm HI+H_2}^{1}$ are shown as dashed cyan, dashed blue, and dashed green line, respectively.}
\label{fig:comp_SPSPr_HIH2SF}\end{minipage}
\end{figure*}
\begin{figure}
\centering
  \resizebox{0.85\hsize}{!}{\includegraphics{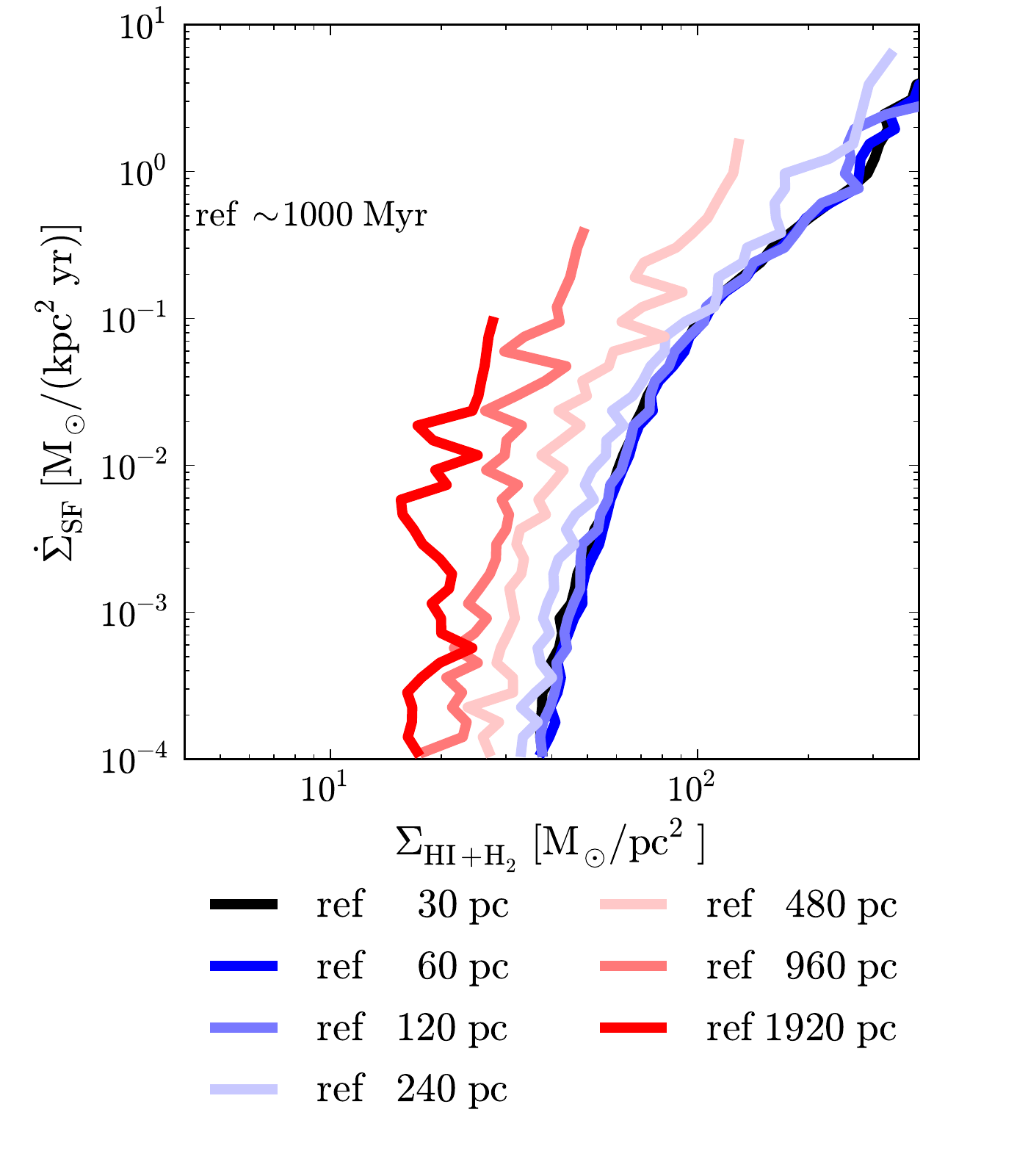}}
\caption{
Star formation column density $\dot{\Sigma}_{\rm SF}$ over $\mathrm{HI+H_2}$ column density $\Sigma_{\rm HI+H_2}$ for different resolution scales $\Delta_{\rm crse}=\left\{30,\ 60,\ 120,\ 240,\ 480,\ 960,\ 1920\right\}\mathrm{pc}$ as solid lines.
The full resolution data from the ref run were coarsened by the appropriate powers of two. 
Additionally, data from 10 consecutive root grid time steps (separated by $\mathrm{d}t\simeq0.65\ \mathrm{Myr}$) around a simulation time of 1 Gyr were averaged to reduce scatter.}
\label{fig:SPr_HIH2SF_res}
\end{figure}
Without following the chemical evolution of molecular and atomic gas, we cannot, unfortunately, distinguish between atomic gas and gas that would observationally be considered molecular. 
So we focus on the combined $\mathrm{HI+H_2}$ content. 
The general shape of $\dot{\Sigma}_{\rm SF}$ over $\mathrm{HI+H_2}$ shown in the left panel of Fig.~\ref{fig:comp_SPSPr_HIH2SF} is similar to observational relations\citep[e.g. in ][]{Schruba2010}, but our distribution is shifted towards higher densities and rates.\\
As demonstrated by \citet{Schruba2010}, the distribution in $\dot{\Sigma}_{\rm SF}$-$\Sigma_{\rm HI+H_2}$ space changes from low to high $\Sigma_{\rm HI+H_2}$ between the vertical 'barrier' behaviour in $\dot{\Sigma}_{\rm SF}$-$\Sigma_{\rm HI}$ space and the linear relation observed for $\mathrm{H_2}$.
This transition happens around the critical density for atomic to molecular conversion, which is metallicity dependent. Increasing the density of gas only leads to an increase in the molecular density, if its density was already above the critical density. 
Because of that the atomic density saturates. This explains the 'vertical barrier'-shaped distribution in $\dot{\Sigma}_{\rm SF}$-$\Sigma_{\rm HI}$ space, if we take into account that the amount of surrounding atomic gas does not affect star formation in the molecular gas.\\
From the high-density tail in the left panel of Fig.~\ref{fig:comp_SPSPr_HIH2SF} one would rather infer a power-law slope $\alpha_{\rm SF,HI+H_2}\simeq1.5$ than a linear relation with $\alpha_{\rm SF,HI+H_2}\simeq1$, but also a power-law slope $\alpha_{\rm SF,HI+H_2}\simeq2$ seems possible. 
At this point we cannot distinguish whether the reason for this is either that stellar feedback prevents the high density tail from being populated up to densities at which $\alpha_{\rm SF,HI+H_2}\simeq1$ could be observed, or the super-linear relation also seen in the equilibrium solutions of BS12 is recovered. 
With increasing metallicity of the star-forming gas the critical density of the atomic to molecular transition drops gradually. This effect is demonstrated by the shift towards lower $\Sigma_{\rm HI+H_2}$ with increasing time in the middle panel of Fig.~\ref{fig:comp_SPSPr_HIH2SF}. 
A linear fit to the tail of the $\dot{\Sigma}_{\rm SF}$--$\Sigma_{HI+H_2}$ distribution in ref after 1~Gyr yields a gas depletion time $\tau_{\rm dep,HI+H_2}\simeq 400\ \mathrm{Myr}$, roughly consistent with $\tau_{\rm dep}(0.32\ \mathrm{M_\odot\ pc^{-3}}>\rho_{\rm thr}>0.1\ \mathrm{M_\odot\ pc^{-3}})$ (see Section~\ref{sec:tdep}).
In the ref run roughly the same amount of stars and metals were produced after 1~Gyr as in the nEnB after 0.4~Gyr. Nevertheless the $\dot{\Sigma}_{\rm SF}$--$\Sigma_{HI+H_2}$ relations are very different. 
While the high-density tail indicating the transition from $\mathrm{HI}$ to $\mathrm{H_2}$ becomes shallower and more prominent in the course of the disc evolution, the shape of $\dot{\Sigma}_{\rm SF}$ over $\Sigma_{HI+H_2}$ in the nEnB run at 0.4~Gyr is about the same as in the early stages of ref. 
This is a consequence of inefficient mixing of the hot, metal-rich material from SNe with cold dense, but still metal-poor material that possibly could be turned into stars. 
The shift between the $\dot{\Sigma}_{\rm SF}$--$\Sigma_{HI+H_2}$ relations for different runs in the right panel of Fig.~\ref{fig:comp_SPSPr_HIH2SF} is mainly caused by gas consumption, as the produced metals have not been mixed into the star-forming material yet.
So, the shift between our results and observational findings is, on the one hand, a consequence of the lower metallicity and higher gas contents, compared to observed local galaxies.\\

On the other hand the resolution scale $\Delta_{\rm crse}$, at which the relation between star formation and gas density is evaluated, influences the critical density of the atomic to molecular transition too, as demonstrated in Fig.~\ref{fig:SPr_HIH2SF_res}. 
Once the averaging scale is considerably larger than a typical star-forming region, i.e. above ~120 pc in the ref run, the transition range in $\dot{\Sigma}_{\rm SF}$--$\Sigma_{\rm HI+H_2}$ space seems to be shifted towards lower densities.
Observational data like in \citet{Bigiel2011} usually correspond to averages over much larger areas, as observations used for statistical analysis of star formation in disc galaxies have spatial resolutions about some $10^2\ \mathrm{pc}$, compared to our $\sim30\ \mathrm{pc}$.
The star formation rate of a given active region is put into relation with a larger volume that contains besides the dense star-forming gas large amounts of ambient, inactive gas. 
Since the tracers have a certain lifetime, observationally measured star formation rates are temporal averages as well, while our data follow the instantaneous star formation rate. However, the shift of the position of the knee in the $\dot{\Sigma}_{\rm SF}$--$\Sigma_{HI+H_2}$ space with coarsening is a combination of resolution effects and the already discussed effects of stellar to gaseous mass ratio and the metallicity in the dense gas of the discs. 
Within a resolution element the fraction of gas that is not directly involved in the star formation process increases with decreasing resolution due to averaging. 
The high density tail in $\dot{\Sigma}_{\rm SF}$-$\Sigma_{HI+H_2}$ space indicating a correlation of star formation with gas density becomes therefore less pronounced in case of coarser resolution.\\
\subsubsection{Turbulence}
\begin{figure*}\begin{minipage}{177mm}
\centering
  \resizebox{0.95\hsize}{!}{\includegraphics{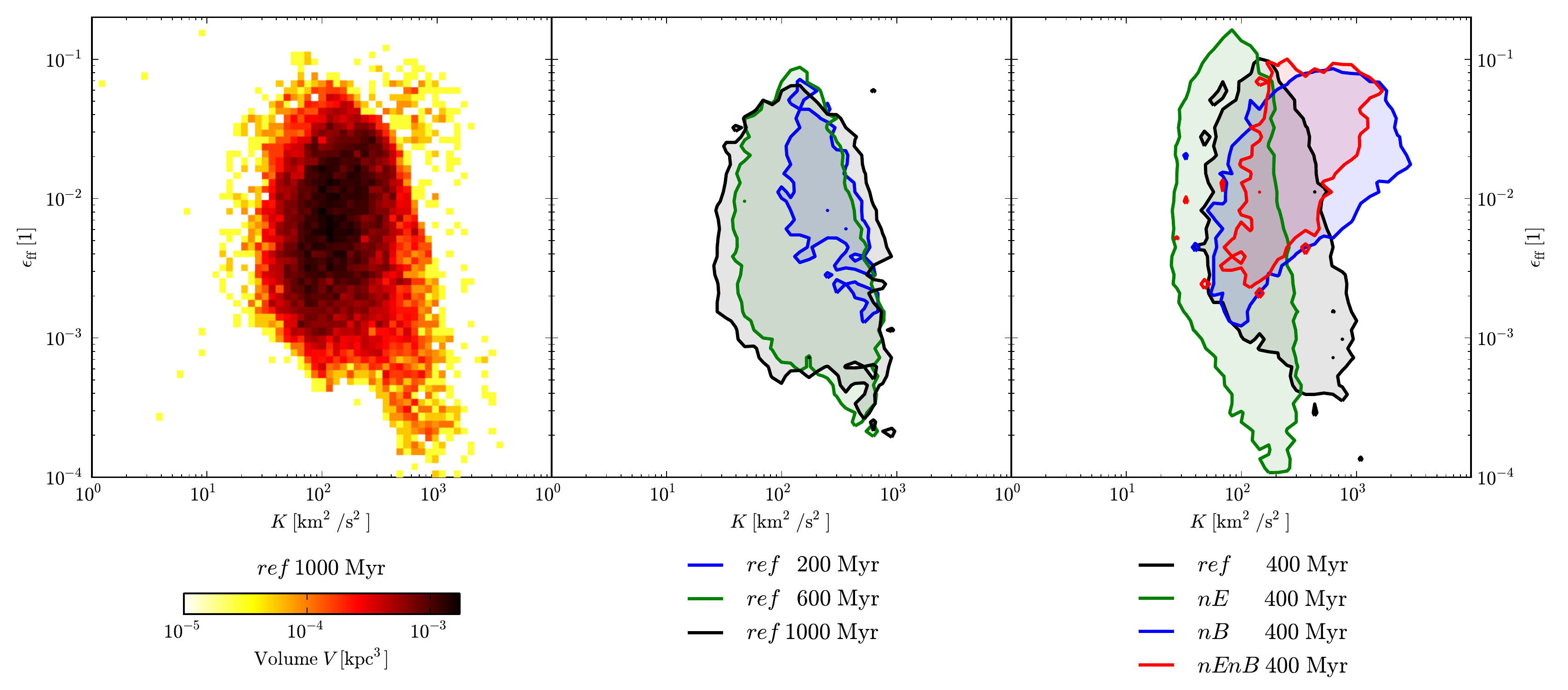}}
\caption{Star formation efficiency $\epsilon_{\rm ff}$ over $K$. Left panel: phase plot for the central region of the disc in the ref run. Middle and right panels: occupied volume $V_{\rm bin}> 3\times 10^{-4}\ \mathrm{kpc^3}$ contours of phase diagrams at different times in the ref run, and at 0.4~Gyr in the different runs. Plots and colours are arranged as in Fig.~\ref{fig:cPPr_SF_Eff}.}
\label{fig:cPC_K_Eff}\end{minipage}
\end{figure*}
The main factor that determines the star formation rate $\dot{\rho}_{\rm SF}$ is the estimated amount of shielded $\mathrm{H_2}$. 
The molecular fraction of the cold phase $f_{\rm H_2}$ is computed by finding the location of the photo-dissociation front in a spherical clump of size $l_{\rm c}$ and density $\rho_{\rm c,pa}$. 
The turbulent state has a major impact on these quantities via two competing effects. 
First, turbulent motions boost the production rate of $\mathrm{H_2}$ via local density enhancements of the cold phase. 
Secondly the turbulent contribution dominates the effective pressure in the cold phase and may play a significant role in the warm-phase pressure as well. 
For higher $K$ the difference between the phase densities $\rho_{\rm c,pa}\geq\rho$ and $\rho_{\rm w,pa}\leq\rho$ becomes smaller, which partially counteracts the boost of the production rate. 
The interplay between these processes determines the minimum $K$ for star formation. 
In the left panel of Fig.~\ref{fig:cPC_K_Eff} one can see that in the ref run star formation is strongly suppressed for $K>10^3\ \mathrm{km^2\ s^{-2}}$. 
For high $K$ also the impact of feedback becomes important. 
Since feedback causes the gas to expand, however, there is very little high-density gas that is also strongly turbulent. 
This explains the relatively narrow range around $K\simeq100\ \mathrm{km^2s^{-2}}$ in which star formation actually occurs, which corresponds to a velocity dispersion around $\simeq10\ \mathrm{km\ s}^{-1}$ or a turbulent RMS Mach number of about 10, consistent with observations \citep{Shetty12}.\\

The enrichment of the gas with metals lowers the minimum $K$ for which gas of a given density can become molecular. 
This effect can be seen in the middle panel of Fig.~\ref{fig:cPC_K_Eff}. 
The degree of enrichment differs throughout the disc, and hence, star formation is possible for a broader range of $K$.\\

The evolution of an individual star-forming region is easily understood by following its path in the $\epsilon_{\rm ff}$--$K$ diagram. 
Starting at the minimal $K$ for star formation, accretion, collapse, Lyman-feedback, and phase separation rapidly produce turbulence until the production is balanced by dissipation. 
Simultaneously $\dot{\rho}_{\rm SF}$ and $\epsilon_{\rm ff}$ increase and eventually reach a self-regulated state. 
This regime is associated with the most densely populated area in $\epsilon_{\rm ff}$-$K$ space around $K\simeq10^2\ \mathrm{km^2\ s^{-2}}$. 
But once SNe begin to dominate the thermal evolution locally, density, $\epsilon_{\rm ff}$, and $\dot{\rho}_{\rm SF}$ reached their peak values and subsequently decline.
Comparing the contours of the runs with $\epsilon_{\rm SN}=0$ to those with $\epsilon_{\rm  SN}=0.085$ in Fig.~\ref{fig:cPC_K_Eff}, it appears that stars are formed in more turbulent environments in the latter case. 
This reflects the enhancement of $K$ by SN feedback.
Switching the hot phase treatment off reduces the impact of thermal feedback, as all thermal SN feedback energy is instantly mixed into the dense warm gas. 
Since much stronger feedback is needed to dilute the gas, the gaseous disc is clumpier and the clumps tend to be denser in this case. 
This increases both the star formation efficiency and $K$ as shown in the right panel of Fig.~\ref{fig:cPC_K_Eff}. 
Once the feedback begins to dominate in a clump, the subsequent expansion is much faster without hot phase treatment, and hence, star formation with low $\epsilon_{\rm ff}$ in strongly turbulent gas is cut off.
\subsection{Drivers of turbulence}
\begin{figure}
\centering
  \resizebox{0.75\hsize}{!}{\includegraphics{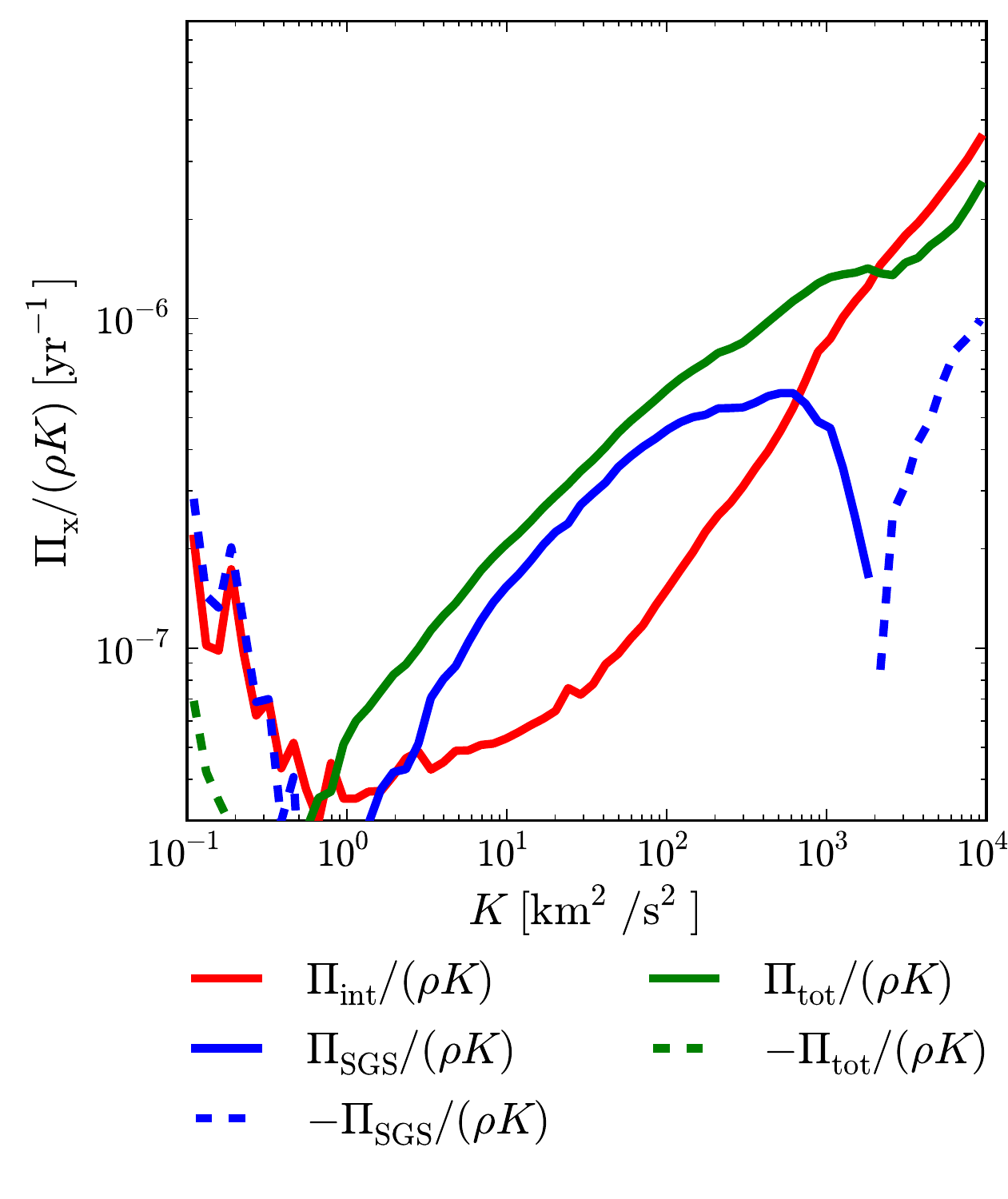}}
\caption{Volume-weighted mean of inverse production time-scales of turbulent sub-grid energy $\Pi/(\rho K)$ over specific unresolved turbulent energy $K$ in the star-forming part of the disc from the ref run after 1 Gyr. The inverse time-scale $\Pi_{\rm SGS}/(\rho K)$ of production by resolved motions via the turbulent stress tensor is shown in blue. The inverse time-scale $\Pi_{\rm int}/(\rho K):=(\Pi_{\rm TI}+\Pi_{\rm SN})/(\rho K)$ of the non-adiabatic MIST-sources is printed in red, and the inverse time-scale $\Pi_{\rm tot}/(\rho K):=(\Pi_{\rm int}+\Pi_{\rm SGS})/(\rho K)$ of the total small scale turbulent energy production. A solid line indicates positive, and a dashed line negative values.}
\label{fig:sPr_K_Pix_ref1000}
\end{figure}
\begin{figure}
\centering
  \resizebox{0.75\hsize}{!}{\includegraphics{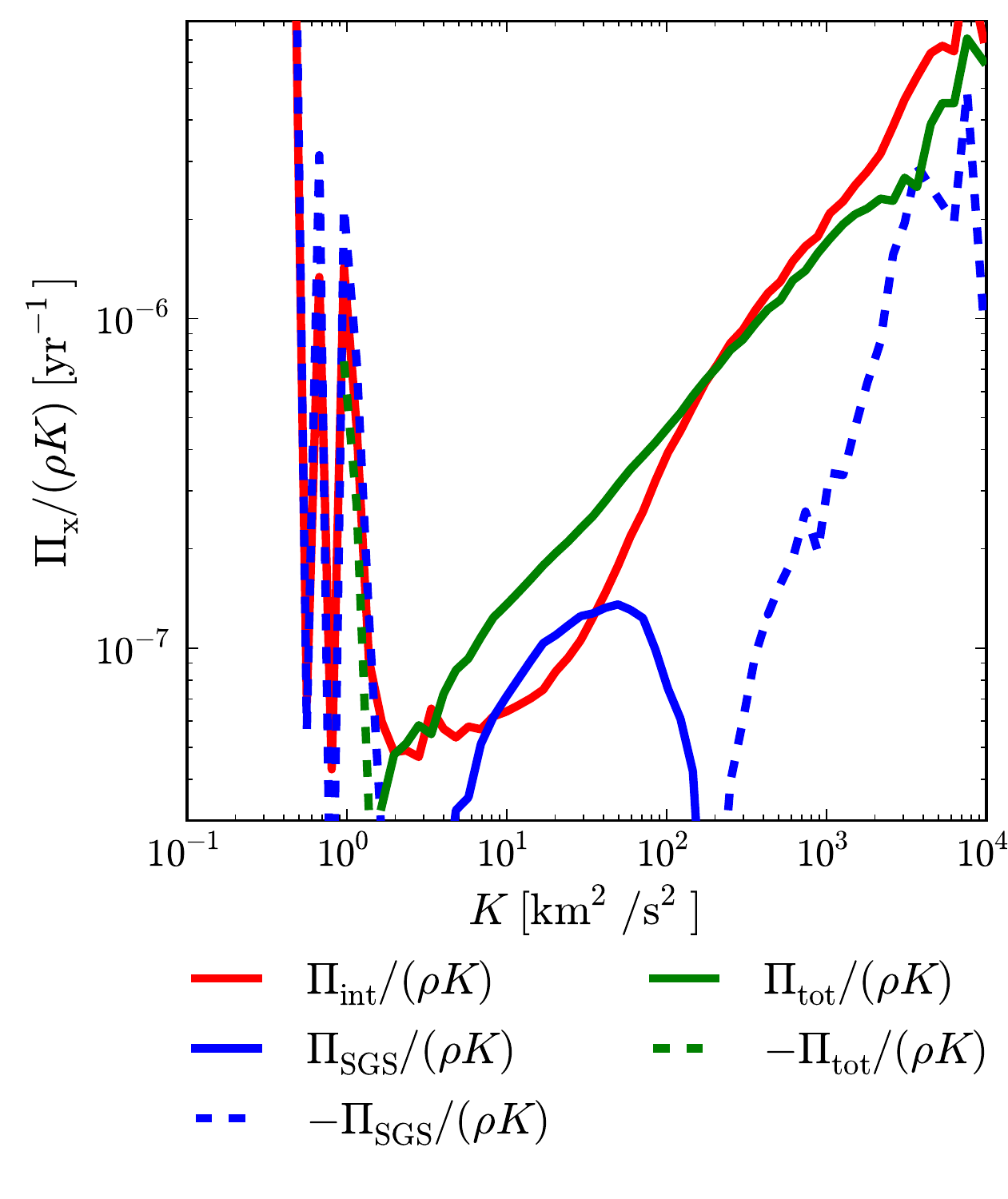}}
\caption{Volume weighted mean of inverse production time-scales of turbulent sub-grid energy $\Pi/(\rho K)$ over specific unresolved turbulent energy $K$ in dense ($\rho>0.032\ \mathrm{M_\odot\ pc^{-3}}$) areas of the disc from the ref run after 1 Gyr. Line colours and styles are arranged as in Fig.~\ref{fig:sPr_K_Pix_ref1000}.}
\label{fig:sPr_K_Pix_ref1000_HD}
\end{figure}
\begin{figure} 
\centering
  \resizebox{0.75\hsize}{!}{\includegraphics{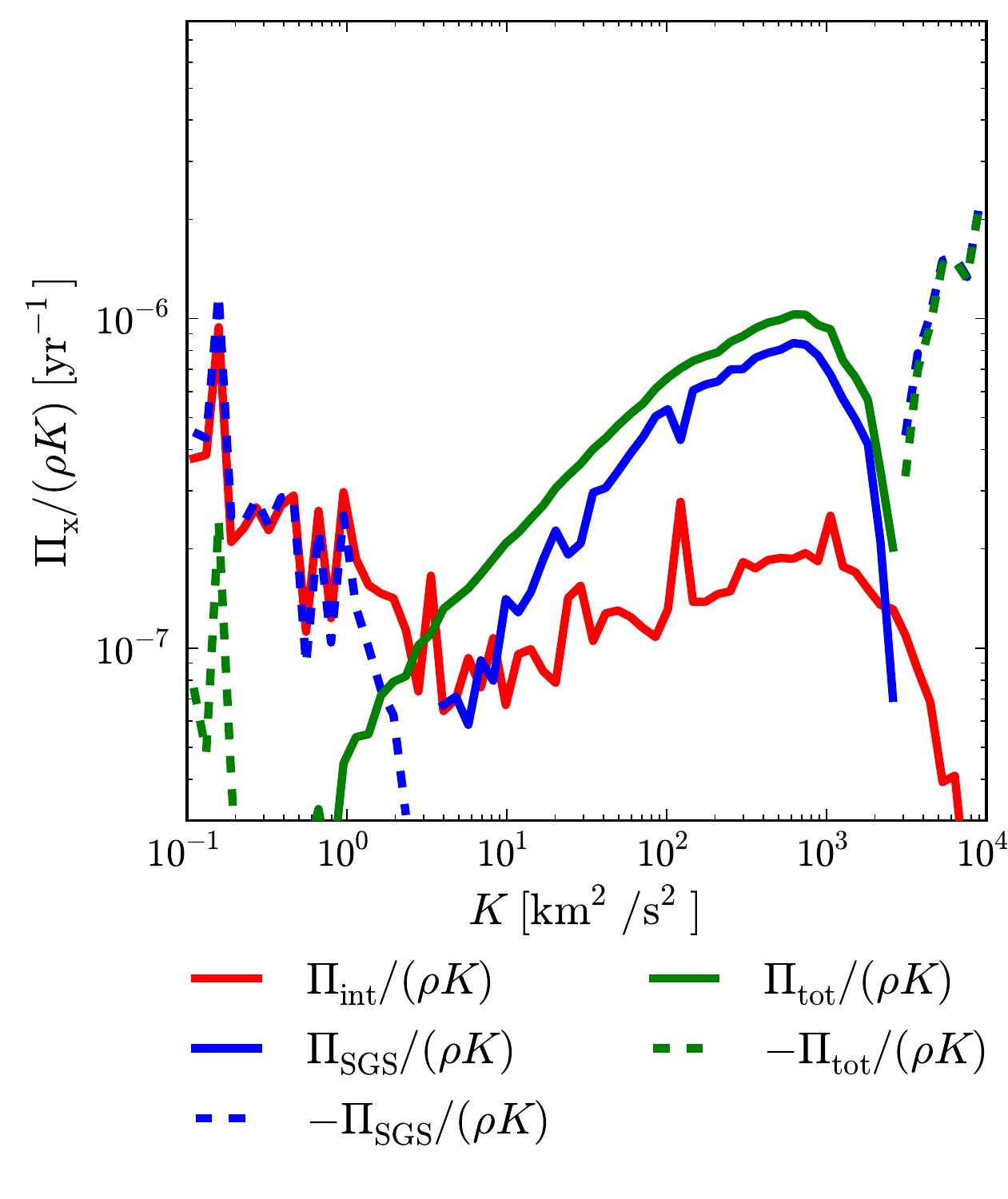}}
\caption{Volume weighted mean of inverse production time-scales of turbulent sub-grid energy $\Pi/(\rho K)$ over specific unresolved turbulent energy $K$ in the star-forming part of the disc from the nE run after 1 Gyr. Line colours and styles are arranged as in Fig.~\ref{fig:sPr_K_Pix_ref1000}.}
\label{fig:sPr_K_Pix_nE1000}
\end{figure}
In the MIST model implementation we consider two groups of sources of SGS turbulence. First there is the production of small-scale turbulence by large-scale shear and compression. The corresponding source term $\Pi_{\rm SGS}$ in equation~(\ref{eq:pi_casc}) may be both positive or negative, indicating whether small-scale turbulence is driven by resolved motions (direct cascade) or the other way around (inverse cascade). The internal sources 
\begin{equation}
 \Pi_{\rm int} := \Pi_{\rm SN} + \Pi_{\rm TI}
\end{equation}
are specific to MIST. The physical processes modelled here are the phase separation due to thermal instability ($\Pi_{\rm TI}$, equation~(\ref{eq:pi_ti})), and small-scale motions caused by SNe bubbles and instabilities in their blast waves ($\Pi_{\rm SN}$, equation~(\ref{eq:pi_sn})). Our model enables us to disentangle the contributions to the total production rate $\Pi_{\rm tot}:=\Pi_{\rm SGS}+\Pi_{\rm int}$ from the turbulent cascade and internal sources to answer the question, which processes are most relevant in which regime.\\

We can distinguish four regimes, as demonstrated in Fig.~\ref{fig:sPr_K_Pix_ref1000} for the ref run:
\begin{enumerate}
 \item $K\lesssim10\ \mathrm{km^2\ s^{-2}}$: negative $\Pi_{\rm SGS}$ indicate expanding environments. In this case, it is the signature of infalling dilute gas. The material found in this regime is most likely to fall freely towards some dense clump. 
$\Pi_{\rm int}\lesssim-\Pi_{\rm SGS}$ is caused by the enforcement of the floor $K_{\rm min}=0.05\ \mathrm{km^2\ s^{-2}}\ll e_{\rm c}$.\footnote{A minimum level of turbulence is needed for the implementation of the SF11-model to work properly, as $\Pi_{\rm SGS}$ depends directly on the instant value of $K$ (see Eqn.~(\ref{eq:pi_casc},\ref{eq:turb_stresses})).}
 \item $10\ \mathrm{km^2\ s^{-2}}\lesssim K\lesssim100\ \mathrm{km^2\ s^{-2}}$: turbulence is mainly supported through the turbulent cascade $\Pi_{\rm SGS}$. A sub-dominant contribution is $\Pi_{\rm TI}$. This regime is typical for star-forming clumps in their early evolutionary stages.
 \item $100\ \mathrm{km^2\ s^{-2}}\lesssim K\lesssim1000\ \mathrm{km^2\ s^{-2}}$: $\Pi_{\rm SGS}$ is the main driver of unresolved turbulence in low density environments, but $\Pi_{\rm int}$ dominates and $\Pi_{\rm SGS}$ becomes negative in dense environments (see Fig.~\ref{fig:sPr_K_Pix_ref1000_HD}). The latter case indicates an expanding environment. For greater $K$ the internal sources $\Pi_{\rm int}$ become increasingly dominated by $\Pi_{\rm SN}$. Star-forming regions in late evolutionary stages reside in this regime.
 \item $K\gtrsim1000\ \mathrm{km^2\ s^{-2}}$: SNe feedback dominates not only the thermal evolution but also all other contributions to turbulence production. The gas in this regime is hot and rapidly expanding.
\end{enumerate}
The power-law behaviour $\Pi_{\rm tot}/(\rho K)\propto K^{\sim0.5}$ for intermediate $K$ indicates that $\Pi_{\rm tot}$ is, on average, balanced by turbulent dissipation $\rho\epsilon_{\rm SGS}\propto\rho K^{1.5}$.
While $\Pi_{\rm TI}$ is slightly enhanced in the absence of $\Pi_{\rm SN}$ (i.e. in the nE and nEnB run), $\Pi_{\rm SGS}$ remains roughly unchanged (see Fig.~\ref{fig:sPr_K_Pix_nE1000}). In this case the low $K$-regime is extended up to $K\lesssim50\ \mathrm{km^2\ s^{-2}}$. 
A further consequence of $\Pi_{\rm SN}=0$ is the lack of effective turbulence production for $K > 10^3\ \mathrm{km^2\ s^{-2}}$.
\subsection{Resolution study}\label{sec:resolution}
\begin{figure}
\centering
  \resizebox{1.0\hsize}{!}{\includegraphics{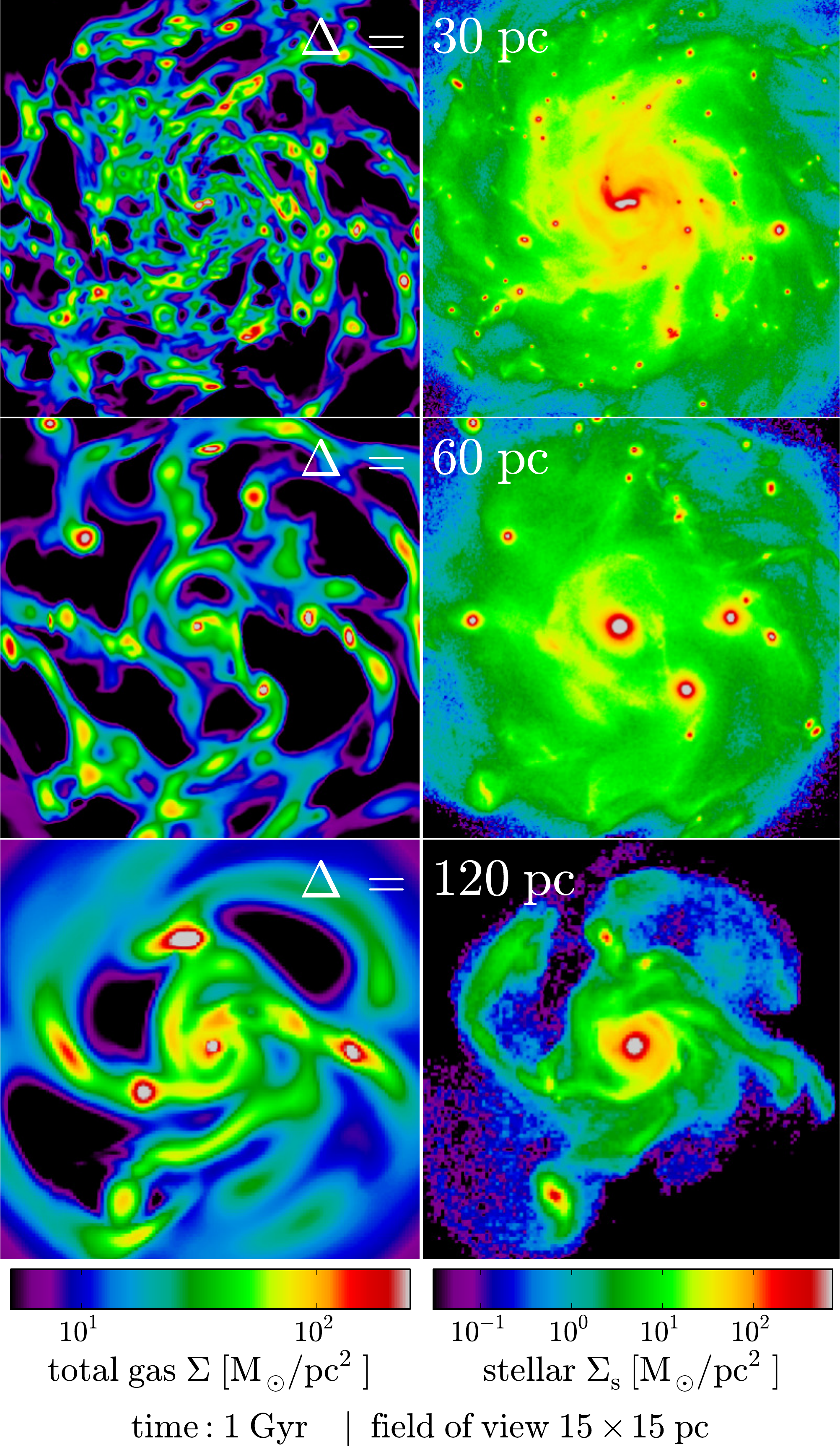}}
\caption{Comparison of total gas surface density $\Sigma$ in the left column and the stellar surface density $\Sigma_{\rm s}$ in the right column between the runs of different effective resolution (from top to bottom: ref, lres5, lres4) 1~Gyr after start of simulation.}
\label{fig:runs_synopsis_res}
\end{figure}
\begin{figure}
\centering
  \resizebox{1.0\hsize}{!}{\includegraphics{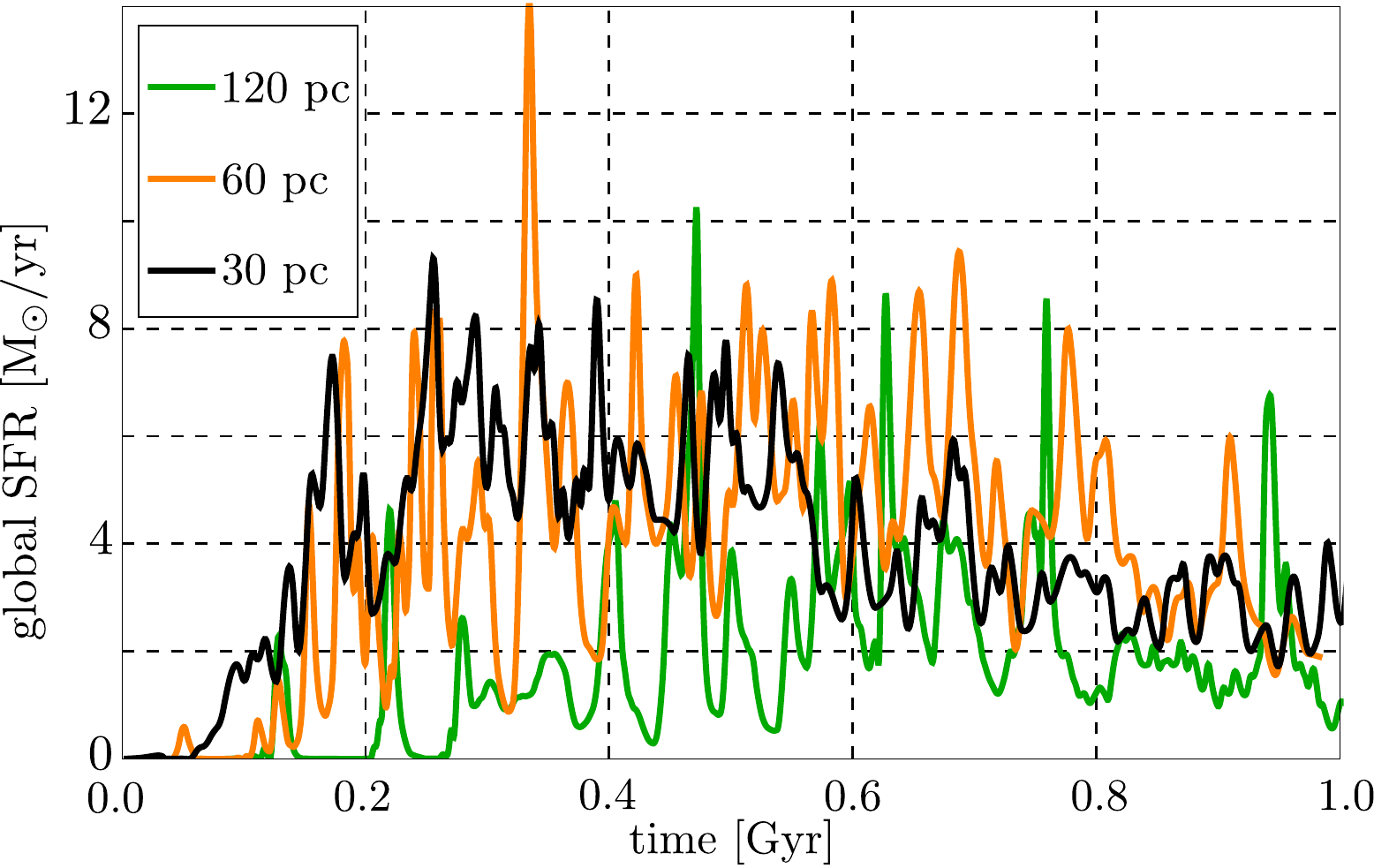}}
\caption{Global star formation rate $\dot{M}_{\rm SF}$ over simulation time for the different runs lres4, lres5, and ref in green, orange, and black, respectively.}
\label{fig:globalSFR_res}
\end{figure}
With decreasing resolution less of the dynamics in the discs is resolved.
In particular the relative importance of the stellar feedback compared to gravity is reduced, as the feedback energy is deposited in a larger volume. This leads to more clumpy discs in the low-resolution runs (see Fig.~\ref{fig:runs_synopsis_res}). 
The resulting gaseous structures like knots and connecting features are much more extended and more massive than one would expect from the ratio of resolutions alone. 
BS12 have shown that the size of the reference volume, which corresponds to a numerical resolution element, does not affect the equilibrium solutions. 
However, the time-scales related to an evolution from an arbitrary state towards equilibrium in the BS12 model are considerably longer for larger $\Delta$, which is caused by additional turbulent modes in a larger reference volume.\\
This effect could be compensated for by adjusting $\epsilon_{\rm SN}$, $e_{\rm h}$, and $\tau_{\rm h}$. 
The physical reasoning is that an increased fraction of the energy released by SNe is present in the form of motions on larger spatial scales driven by the expanding SNe bubbles which provide additional pressure support.\\
Despite the differences in the disc structure, the global star formation rates in the ref and lres5 run are in good agreement with each other (see Fig.~\ref{fig:globalSFR_res}), although the initial transient phase in the lres5 run ($\Delta\simeq60\ \mathrm{pc}$) lasts longer than in the ref run ($\Delta\simeq30\ \mathrm{pc}$). In both runs about $\sim$30 per cent of the initial mass were transformed into stars after 1~Gyr. 
Star formation in the lres4 run ($\Delta\simeq120\ \mathrm{pc}$) is less efficient, such that only $\sim$15 per cent of the mass was turned into stars after 1~Gyr. 
Due to lack of resolution and the effect discussed above, the relevant structures for star formation either are not sufficiently resolved or simply do not form.\\
\begin{figure}
\centering
  \resizebox{0.85\hsize}{!}{\includegraphics{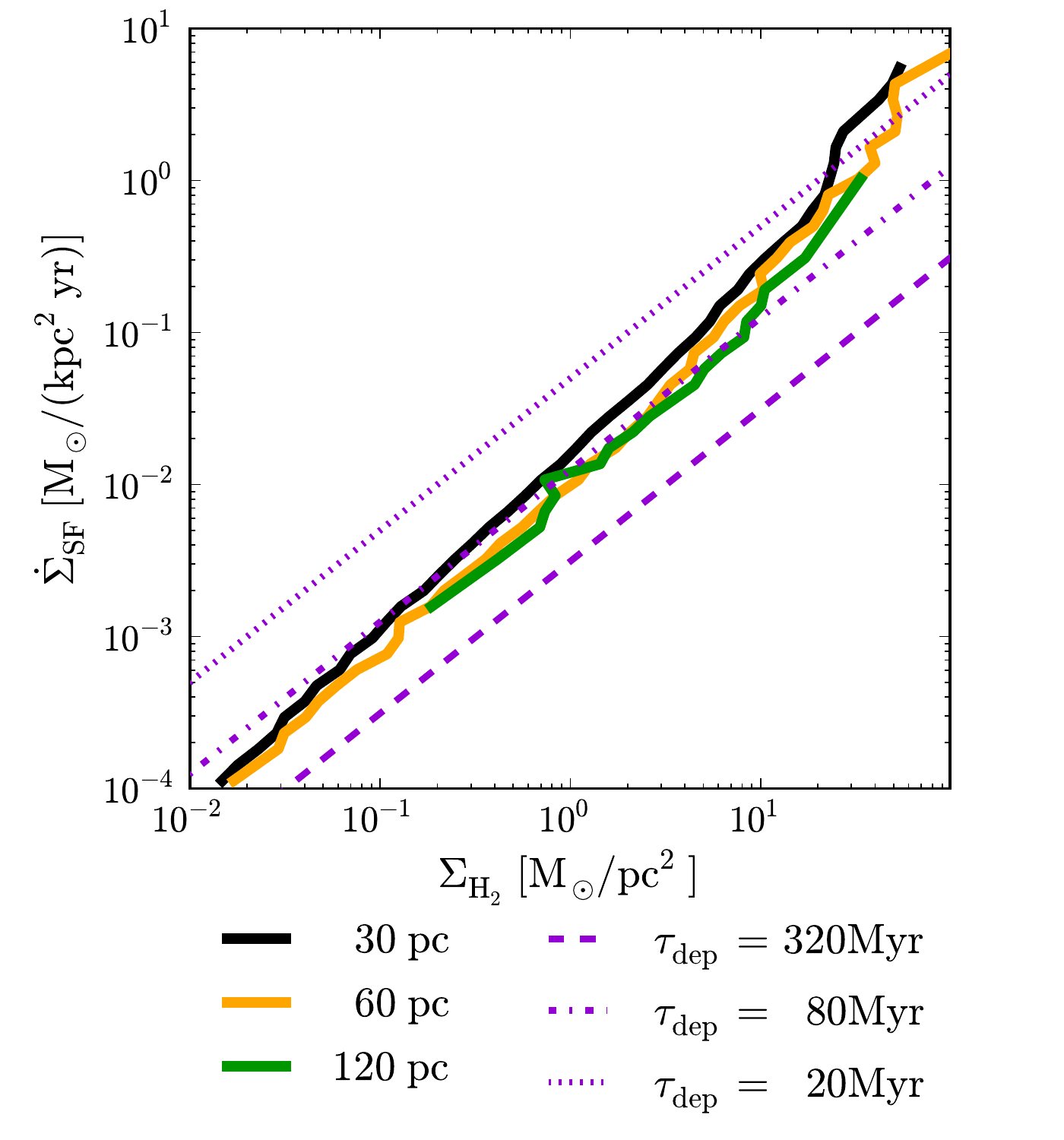}}
\caption{Star formation column density $\dot{\Sigma}_{\rm SF}$ over $\mathrm{H_2}$ column density $\Sigma_{\rm H_2}$ for runs with different numerical resolution: lres4, lres5, and ref in green, orange, and black, respectively.}
\label{fig:SPr_H2SF_resrun}
\end{figure}
\begin{figure}
\centering
  \resizebox{0.85\hsize}{!}{\includegraphics{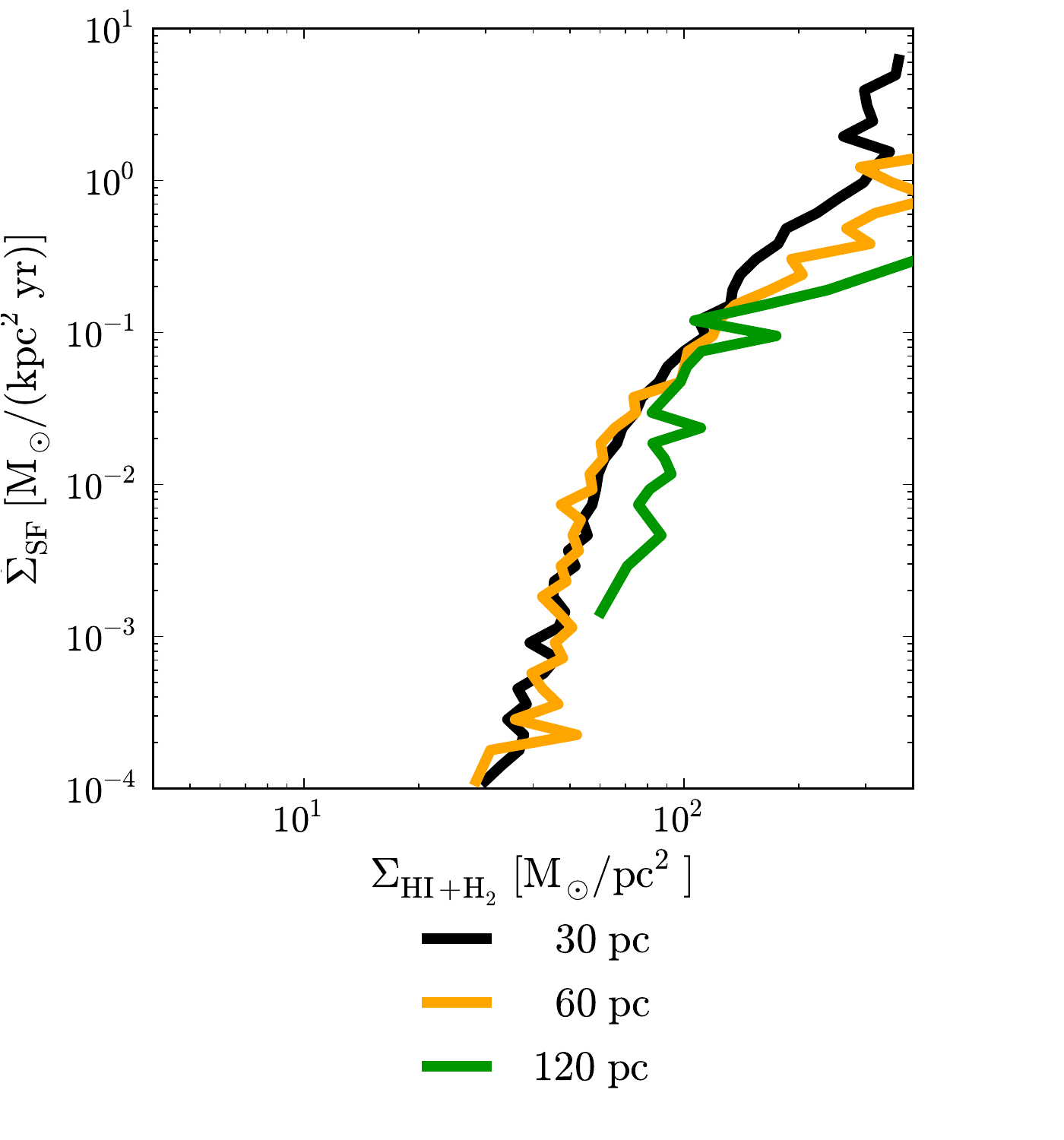}}
\caption{Star formation column density $\dot{\Sigma}_{\rm SF}$ over $\mathrm{HI+H_2}$ column density $\Sigma_{\rm HI+H_2}$ for runs with different numerical resolution: lres4, lres5, and ref in green, orange, and black, respectively.}
\label{fig:SPr_HIH2SF_resrun}
\end{figure}
Regardless of the differences between the runs, the star formation versus density relations, as shown in Fig.~\ref{fig:SPr_H2SF_resrun} for $\dot{\Sigma}_{\rm SF}$ over $\Sigma_{\rm H_2}$ and in Fig.~\ref{fig:SPr_HIH2SF_resrun} for $\dot{\Sigma}_{\rm SF}$ over $\Sigma_{\rm HI+H_2}$, stay roughly unchanged, which is a consequence of the internal regulation of MIST. The slight shift in the case of the plots with $\Delta=120\ \mathrm{pc}$ in Figs~\ref{fig:SPr_HIH2SF_res} and \ref{fig:SPr_HIH2SF_resrun} is caused by the generally lower star formation rate in the lres4 run compared to the ref run.\\
\subsection{Impact of subgrid scale model}\label{sec:simpleISM}
\begin{figure}
\centering
  \resizebox{1.0\hsize}{!}{\includegraphics{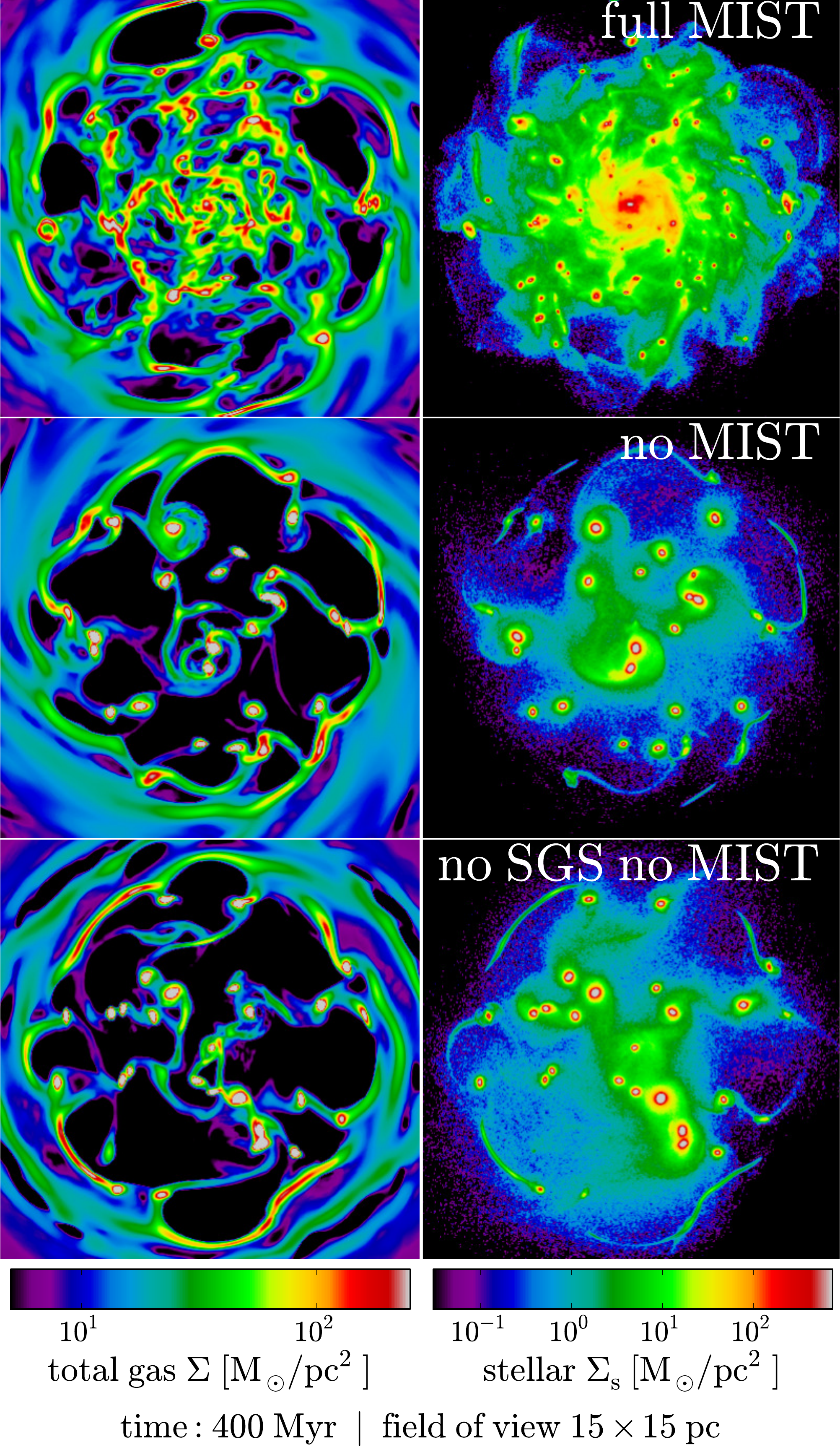}}
\caption{Comparison of total gas surface density $\Sigma$ in the left column and the stellar surface density $\Sigma_{\rm s}$ in the right column between the runs using different ISM models (from top to bottom: ref, sSF, sSF2) 400~Myr after start of simulation.}
\label{fig:runs_synopsis_simpl}
\end{figure}
\begin{figure}
\centering
  \resizebox{0.99\hsize}{!}{\includegraphics{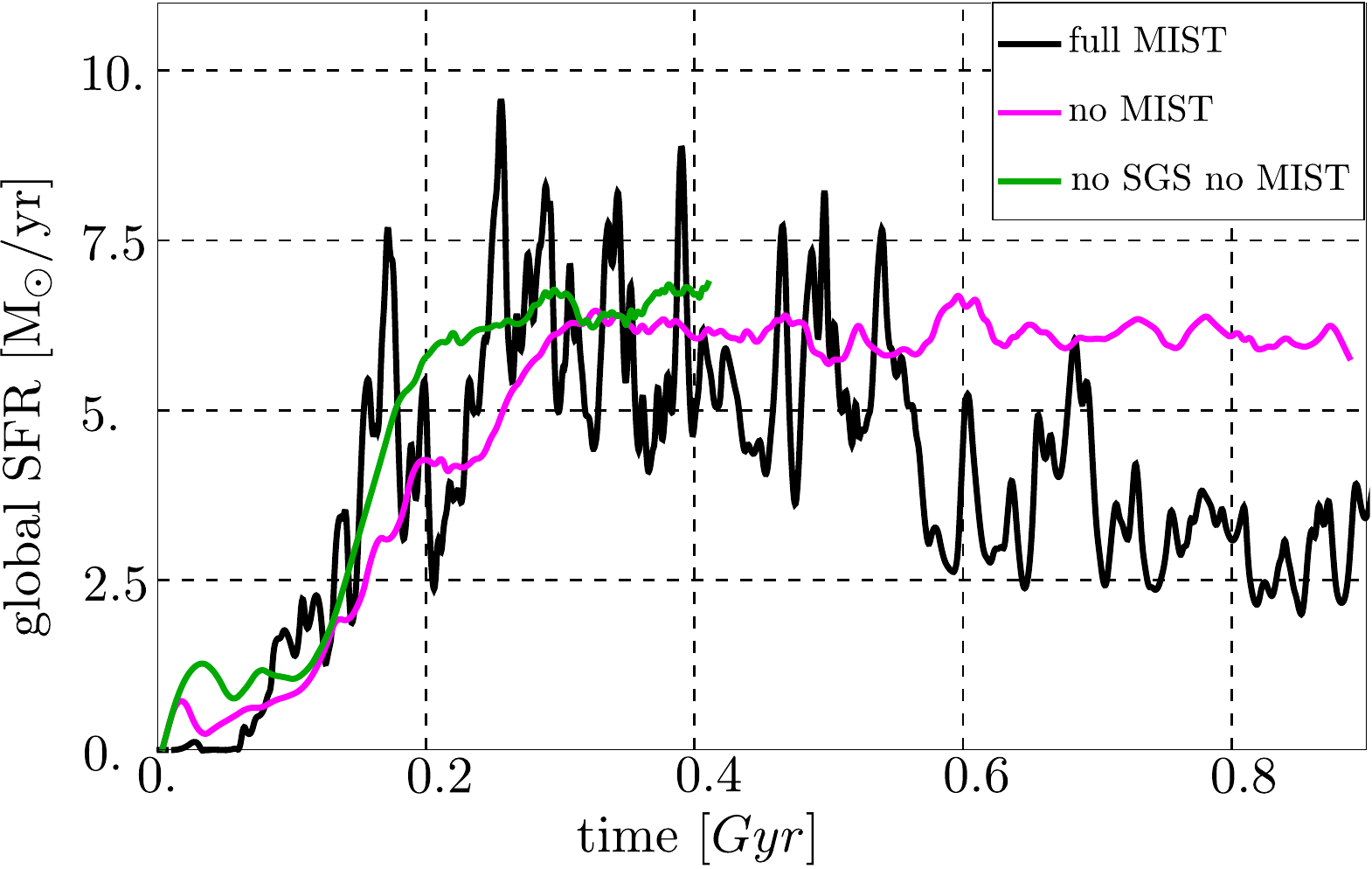}}
\caption{Global star formation rate $\dot{M}_{\rm SF}$ over simulation time for the different runs ref, sSF, and sSF2 in black, purple, and green, respectively.}
\label{fig:globalSFR_simpl}
\end{figure}
\begin{figure}
\centering
  \resizebox{0.85\hsize}{!}{\includegraphics{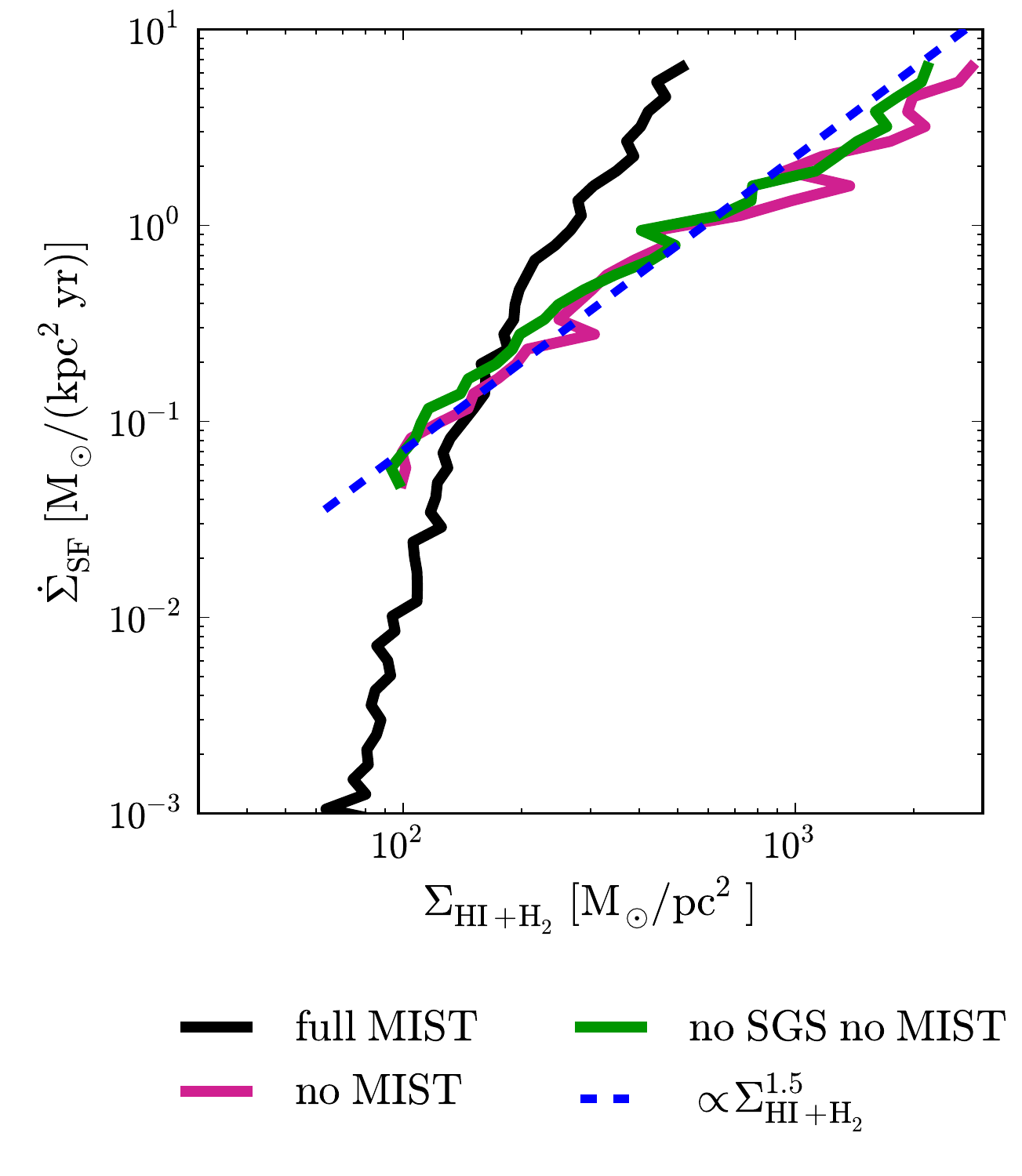}}
\caption{Star formation column density $\dot{\Sigma}_{\rm SF}$ over $\mathrm{HI+H_2}$ column density $\Sigma_{\rm HI+H_2}$ for the runs ref, sSF, and sSF2 in black, purple, and green, respectively.}
\label{fig:SPr_HIH2SF_simpl}
\end{figure}

Fig.~\ref{fig:runs_synopsis_simpl} shows that the stellar discs in the runs without MIST (sSF and sSF2) are dominated by a few very massive stellar clusters after 400 Myr. 
The gas follows the distribution of stars and is consequently concentrated in the stellar clusters. 
Star formation occurs only in those places, since there is no high-density gas elsewhere. Clearly, the \citet{Truelove1997} criterion is not fulfilled within those clusters, as neither SGS turbulence energy nor SNe energy feedback can support the gas sufficiently in both runs. 
The SGS model is necessary to keep the gas (and in the consequence the stars) from clustering too much in order to obtain a realistically smooth and flocculent disc.\\
The global star formation rate of a few $\mathrm{M}_\odot\ \mathrm{yr^{-1}}$ in sSF and sSF2 is comparable to the ref run (see Fig.~\ref{fig:globalSFR_simpl}). 
This is a consequence of the choice of the parameters in the simplified star formation model ($\varepsilon_{\rm sSF}=0.01$, $\rho_{\rm sSF,min}=50m_{\rm H}\ cm^{-3}$ and $T_{\rm sSF,max}=1.5\times10^4\ \mathrm{K}$), which are chosen to match the average star formation properties in the runs with MIST. 
However, both runs with the simplified model lack the strong variations on time-scales between 10 and 30~Myr seen in runs with MIST. 
As pointed out in Section~\ref{sec:refrun} and \ref{sec:tdep}, the variations are related to the life cycle of individual star-forming regions. 
The SNe feedback disrupting the cold dense gas after a phase of intense star formation limits the lifetime of an active region in the MIST runs.
The lifetime of these regions is not limited in the non-MIST runs, as the feedback energy is radiated away before it could affect the gas. 
This is known as the so-called 'over-cooling' problem.
Moreover, the stellar particles are not inserted with peculiar velocities, representing the unresolved motions of the star-forming gas (see Section~\ref{sec:spart_creation}).
These differences in the treatment of star formation and feedback result in the formation of large strongly bound clusters, in which continuous star formation is fueled by the accumulation of gas through accretion and mergers into even larger clusters. 
This is a runaway process.\\
After an initial transient phase the global star formation rate reaches a plateau around $6\ \mathrm{M}_\odot\ \mathrm{yr^{-1}}$ (see Fig.~\ref{fig:globalSFR_simpl}). 
Due to the stabilizing effect of the SGS turbulence energy the transient phase in sSF lasts 50~Myr longer than in sSF2. 
The gradual decline of the star formation rate seen in runs with MIST due to metal enrichment (see Section~\ref{sec:tdep}) does not occur in sSF and sSF2, as neither $\varepsilon_{\rm sSF}$ nor $\rho_{\rm sSF,min}$ depend on metals. 
The KS relation, depicted in Fig.~\ref{fig:SPr_HIH2SF_simpl}, shows a tight correlation with the expected slope of $\dot{\Sigma}_{\rm SF}\propto\Sigma^{1.5}$.
\section{Discussion and Conclusions}\label{sec:conclude}
In this paper we introduced MIST, which is a SGS model based on the semi-analytical BS12 model \citep{BS12} for the turbulent multi-phase ISM. 
We implemented MIST into the code \textsc{Nyx} \citep{NYX} to run adaptively refined LES of IDG with different stellar feedback parameters. 
For the first time, a complete treatment of the numerically unresolved turbulence energy via the SGS model of SF11 is applied in such simulations.
The star formation recipe follows \citet{Krumholz2009} and \citet{PadNord09}. In our fiducial galaxy model, supernova feedback produces both SGS turbulence energy and heat. 
Since the injected heat would instantly cool away in dense environments, we suppress cooling in the hot gas produced by feedback over a time-scale that is given by the typical size and expansion velocities of hot SN bubbles in the ISM. 
Comparison runs demonstrate that the following properties of star formation in disc galaxies are reproduced
if both turbulent and thermal feedback are applied: 
\begin{enumerate}
\item an average total gas star formation efficiency $\epsilon_{\rm ff}\simeq0.01$
\item $\epsilon_{\rm ff}$ is enhanced in dense environments
\item a galactic depletion time-scale $\tau_{\rm dep}\sim0.3\ldots1\ \mathrm{Gyr}$ in a gas-rich galaxy
\item a linear relationship between $\dot{\Sigma}_{\rm SF}$ and $\Sigma_{\rm H_2}$
\item a lifespan of molecular clouds $t_{\rm GMC}\sim10\ldots30\ \mathrm{Myr}$
\item a velocity dispersion in star-forming regions $\sigma_{\rm SF}\simeq10\ \mathrm{km\ s^{-1}}$
\end{enumerate}
We observe three modes of star formation in our simulations. First, stars are formed in the metal-poor gas that occurs in isolated clouds. In this material, the threshold density for star formation is rather high. 
Therefore, the gas is quickly turned into stars and the ensuing feedback disrupts the cloud violently in the simulation with full feedback. 
Subsequent dispersion of the gas effectively quenches star formation, as the expanding shell is diluted below the threshold density. 
Typically, a stellar cluster is left behind that survives longer than the original gas cloud. 
Eventually, the stars are dispersed in the disc's potential or merge into more stable and massive clusters. 
With increasing amounts of metals in the gas, a different mode of star formation emerges. Owing to the decreasing threshold density, active star formation is possible in more extended regions of lower density. 
Although star formation is still quenched by feedback in these regions, waves of star formation propagate through the metal-rich inner parts of the disc. Since the feedback is less violent, the winds launched in this mode are slower. 
The third mode of star formation is hosted by massive stellar clusters that locally dominate the gravitational potential. 
They are formed via mergers of the clusters generated in metal-poor star-forming regions. 
Once they are massive enough, they begin to accrete gas from their surroundings, and consequently host star formation. 
Depending on their mass and the chosen feedback parameters, star formation is continuous or intermittent. 
The massive clusters are candidates for globular cluster progenitors, although they continue to grow and slowly spiral towards the center in our isolated disc simulation, where they merge into the central stellar agglomeration. 
In a cosmological galaxy simulation, mergers could interrupt this process by kicking the clusters out of the disc plane and thereby cutting them off of gas supply. 
Moreover, it is possible that late feedback from SNe of type Ia \citep{Agertz2013} has an impact on the star formation induced by clusters. 
In the current implementation, stellar populations older than roughly 40~Myr are quiescent. Taking SNe~Ia into account, the period of active feedback might increase to a few 100~Myr. 
Although this type of feedback is much less intense than the feedback caused by SNe~II, it might prevent the accretion of gas in dense stellar clusters. Also so-called 'early feedback' \citep[e.g. see][]{Stinson2013} in the form of turbulent feedback due to stellar winds from massive stars might make residual stellar clusters more prone to disruption, since the velocity dispersion of the stars would be enhanced.

For all of our runs using MIST, we found a robust, almost linear relationship between shielded molecular gas and the star formation rate, which is a consequence of self-regulatory processes in MIST. 
This kind of relation also appears in the semi-analytic model (BS12) and in observations \citep{Bigiel2011}. 
The related $\mathrm{H_2}$-consumption time-scale of 40 to 80~Myr is in good agreement with observations of actively star-forming clouds \citep{Murray2011}. 
The global gas depletion time-scale of about 0.5-1~Gyr, which is inferred from the mean star formation rate in the simulation, is in agreement with the observed gas depletion in gas-rich galaxies \citep{Daddi2010}. 
The gradual increase of this time-scale suggests that depletion times of a few Gyr
could be reached in later evolutionary stages, comparable to those observed for local galaxies. The gas consumption time-scale inferred from the relation between $\dot{\Sigma}_{\rm SF}$ and$\Sigma_{\rm HI+H_2}$ is also in agreement with observations of gas-rich galaxies. 
These results confirm our method of computing the star formation rate. However, the amount of molecular gas following from our model is generally too low to be consistent with observations. The missing $\mathrm{H_2}$ is a consequence of using a Str\"omgren-like ansatz to obtain an equilibrium solution instead of following the chemical evolution. 
Although this allows us to reliably estimate the amount of molecular gas that resides in shielded areas, a large fraction of the $\mathrm{H_2}$ mass would be found in surrounding regions if detailed chemical reaction networks were computed in the cold and warm phases.  
As the star formation rates in our simulation reproduce observations reasonably well, it appears that the shielding from radiation, and hence the lack of heating, is the major factor that controls star formation.

While the star formation efficiency $\epsilon_{\rm PN}$ with respect to the shielded gas is almost constant for all star-forming regions, regardless of density and metallicity, the efficiency $\epsilon_{\rm ff}$ with respect to the total gas density does vary due to the impact of metals and density on the shielded $\mathrm{H_2}$ content. 
This results in an enhanced efficiency $\epsilon_{\rm ff}$ in regions with high star formation rate, contrary to the sometimes employed constant efficiency parameters
\citep[e.g.][]{Agertz2013}. 
Since the production of $\mathrm{H_2}$ strongly increases with the density of the unresolved cold-gas clumps, which is enhanced by turbulence via the dependence of the cold-gas density PDF on the turbulent Mach number in our model, a minimal level of SGS turbulence energy is necessary for star formation. 
Stellar feedback, which is associated with very high SGS turbulence energy, reduces star formation. Consequently, star formation occurs only for an intermediate range of SGS turbulence energy, centred around a peak value that roughly corresponds to a velocity dispersion of 10~$\mathrm{km\ s^{-1}}$, comparable to observed values \citet{Shetty12}. 
Apart from feedback, the turbulent cascade is an important source for building up the moderate level of turbulence in star-forming regions before SN feedback becomes dominant.\\
While \citet{Agertz2013} make efforts towards accounting for all relevant feedback mechanisms - these are stellar winds from massive stars, radiation pressure, SNe~II, SNe~Ia, and mass loss by mass-poor stars - in a realistic fashion by taking age and metallicity of a stellar population into account and using appropriate application channels and schemes for each of them reflecting their physical impact on the ISM, their star formation recipe is relatively simple. 
In contrast to their approach, we focus on a more elaborate way to describe the sub-resolution structure and processes in the ISM.
The non-thermal pressure - actually the trace of the turbulent stress tensor - helps to stabilize gas against gravity particularly in cold and dense environments. The SGS-energy allows us to collect the effect of subresolution-scale motions excited by SNe, which is our counterpart to the momentum feedback in \citet{Agertz2013}. 
While over-cooling is usually avoided by suppressing all cooling for period of more than 10~Myr, the combined effect of non-thermal pressure and turbulent feedback reduces the need for delayed cooling in areas of active SNe feedback in our simulations.\\

MIST coupled to an SGS model is suitable for the use in cosmological zoom-in simulations with effective resolutions of about 10 to 100~pc without substantial modifications. 
However, the current model framework of MIST has to be adjusted to treat extremely metal-poor gas and the radiation background consistently.
Apart from that the incorporation of reaction networks to track the actual molecular content in the different gas phases and of additional stellar feedback mechanisms like stellar winds from massive stars and SNe~Ia will further enhance the model.
\section*{Acknowledgements}
HB was financially supported by the CRC 963 of the German Research Council.
The work of AA was supported by the SciDAC FASTMath Institute, funded by the Scientific Discovery through Advanced Computing (SciDAC) program funded by U.S. Department of Energy.
HB, WS, and JCN acknowledge financial support by the German Research Council for visits at LBNL. 
We thank Hsiang-Hsu Wang for discussions on the initial conditions of idealized IDG.
The simulations presented in this article were performed on the SuperMUC of the LRZ (project pr47bi) in Germany. 
We also acknowledge the yt toolkit by \citet{yt} that was used for our analysis of numerical data.
We owe thanks to the referee B. Robertson for a careful and helpful report that helped us to improve this paper.

\bibliography{paper}
\appendix

\bsp

\label{lastpage}

\end{document}